\documentclass[reprint,nofootinbib,aps,superscriptaddress,amsmath,amssymb,onecolumn,11pt]{revtex4-2}

\usepackage{graphicx}
\usepackage{dcolumn}
\usepackage{bm}
\usepackage{braket}
\usepackage{breqn}
\usepackage{mathrsfs}
\usepackage{color}

\usepackage[colorlinks=true,linkcolor=red,citecolor=blue]{hyperref}

\begin{document}

\title{Penrose process in magnetized non-Kerr rotating spacetime with anomalous quadrupole moment}

\author{Shao-Jun Zhang}
\email{sjzhang@zjut.edu.cn}
\affiliation{School of Physics$,$ Zhejiang University of Technology$,$ Hangzhou 310032$,$ China}
\affiliation{Institute for Theoretical Physics and Cosmology$,$ Zhejiang University of Technology$,$ Hangzhou 310032$,$ China}
\date{\today}

\begin{abstract}
We investigate the magnetic Penrose process in the Quevedo-Mashhoon spacetime, immersed in a uniform magnetic field \(B\). This metric is a stationary, axisymmetric, asymptotically flat vacuum solution to Einstein's equations with an arbitrary anomalous quadrupole moment \({\cal Q}\). A non-vanishing \({\cal Q}\) significantly modifies the near-horizon geometry, creating a multi-lobe ergoregion. Both \({\cal Q}\) and \(B\) strongly influence the negative-energy region, which can extend well beyond the ergoregion, enabling the magnetic Penrose process to operate far from the ergoregion. Their combined effects allow energy extraction efficiency \(\eta\) to far exceed that of the mechanical Penrose process. The maximum efficiency undergoes three distinct evolutionary stages as \({\cal Q}\) varies. In the absence of the magnetic field, efficiency is optimized for more negative \({\cal Q}\) (yielding a more oblate spacetime than Kerr). When electromagnetic interactions dominate, efficiency peaks when the infalling fragment's charge and \(B\) share the same sign and \({\cal Q}\) is more positive (producing a more prolate spacetime than Kerr). These findings support the magnetic Penrose process as a theoretical framework for high-energy cosmic phenomena (e.g., extragalactic high-energy radiation) and as a tool to test the Kerr hypothesis.

\end{abstract}


\maketitle
  
\section{Introduction}

In recent years, astronomical observations have achieved significant breakthroughs, including the detection of gravitational waves from binary mergers \cite{LIGOScientific:2016sjg,LIGOScientific:2016vlm} and the imaging of supermassive black hole shadows \cite{EventHorizonTelescope:2019dse,EventHorizonTelescope:2022wkp}. The advancements in observational technology now enable us, either currently or in the near future, to probe physics in strong gravity regime with unprecedented precision, thereby testing general relativity (GR) in this regime and delving deeper into the nature of compact astrophysical objects.

The no-hair theorem is a pivotal prediction among the many predictions of GR. It states that any stationary, asymptotically flat vacuum spacetime with an event horizon and no closed timelike curves outside the horizon must be described by the Kerr metric \cite{Carter:1971zc,Robinson:1975bv,Chrusciel:2008js}. The theorem's validity hinges upon a set of rigorous assumptions, hence it is also referred to as the Kerr hypothesis. Despite indirect observational evidence suggesting that numerous detected supermassive astrophysical objects may conform to Kerr black holes, the current observational precision is inadequate to definitively exclude the possibility of these objects being compact exotic objects governed by non-Kerr metrics. Likewise, the existence of alternative Kerr black hole configurations, such as boson stars \cite{Kesden:2004qx,Berti:2006qt}, gravastars \cite{Chirenti:2008pf}, naked singularities, and "dirty" black holes with external matter (e.g., accretion disks) and hairy black holes, remains a viable prospect. Consequently, it is challenging to search for effective techniques and observational methodologies capable of differentiating these alternatives from Kerr black holes.

One approach to testing the Kerr hypothesis involves considering a universally applicable stationary, axisymmetric, asymptotically flat vacuum spacetime \cite{Ryan:1995wh,Ryan:1997hg,Ryan:1997kh}. Such a spacetime can describe the gravitational field around a central compact object (regardless of the nature of the object), and its metric can be expressed in terms of mass moments and current moments \cite{Geroch:1970cd,Hansen:1974zz}. For a Kerr black hole, its spacetime structure is uniquely determined by the black hole's mass and angular momentum, with higher-order multipole moments being determined by these two fundamental parameters via a simple relation
\begin{align}
    M_\ell + i S_\ell = M \left(i \frac{J}{M}\right)^\ell.
\end{align}
Here, $M$ and $J$ represent the mass and angular momentum of the black hole, respectively. $M_\ell$ and $S_\ell$ denote the mass and current multipole moments of the gravitational field at the $\ell$-th order, with $M_0 = M$ and $S_1 = J$. Consequently, the Kerr hypothesis can be tested by identifying deviations in the higher-order multipole moments of massive compact objects from the Kerr solution. For example, future observations could employ detectors like LISA to analyze gravitational wave signals from extreme mass-ratio inspirals (including their quasi-normal modes) for this purpose \cite{Ryan:1995wh,Hughes:2006pm,Brown:2006pj,Berti:2009kk}. Alternatively, future very-long-baseline interferometry (VLBI) experiments could observe the shadows of supermassive black hole candidates to conduct these tests \cite{Bambi:2010fe,Johannsen:2010ru,Johannsen:2010xs}. 

In the strong gravity regime, many peculiar physical processes and phenomena occur, which can serve as observables for probing the spacetime structure. In recent years, we have witnessed a surge in high-energy astrophysical phenomena, including gamma-ray bursts, fast radio bursts, and relativistic jets. These energetic events are characterized by the release of considerable amounts of energy. It is widely believed that black holes or other compact objects play a central role in these high-energy astrophysical processes. Consequently, a thorough investigation of the energy extraction mechanisms from these compact objects is crucial for a complete understanding of such astrophysical phenomena as well as the nature of these objects.

Black hole thermodynamics shows that extremal Kerr black holes can reach a maximum energy extraction efficiency of $\sim 29\%$ \cite{Christodoulou:1970wf,Bekenstein:1973mi}. In 1969, Penrose proposed the first such mechanism, the Penrose process \cite{Penrose:1971uk}: a particle splits into two inside the ergosphere of a rotating black hole. For extremal Kerr black holes, its maximum efficiency is $\sim 20.7\%$, but drops to $<2\%$ for moderate spins (e.g., $a\leq 0.5M$). Beyond low efficiency, it is unfeasible in reality, as the split fragments need a relative velocity exceeding half the speed of light \cite{Bardeen:1972fi,Wald:1974kya}, making such events extremely rare in astrophysics. 

In the mid-1980s, investigations into matter-electromagnetic interactions revitalized the Penrose process, giving rise to its magnetic counterpart (MPP) \cite{dhurandharEnergyextractionProcessesKerr1984,dhurandharEnergyextractionProcessesKerr1984a,1985ApJ...290...12W,parthasarathyHighEfficiencyPenrose1986,Bhat:1985hpc,Wagh:1989zqa}. An external magnetic field facilitates overcoming the relative velocity threshold: the energy of particles in negative-energy orbits now derives from matter-electromagnetic interactions, thereby eliminating constraints on mechanical velocity. MPP can achieve extremely high efficiency (even exceeding $100\%$); for instance, electrons orbiting a stellar-mass black hole require only a milligauss magnetic field to surpass $100\%$ efficiency \cite{Dadhich:2018gmh}. Unlike the mechanical Penrose process, MPP does not depend on the rapid rotation of black holes and can function effectively even under low-spin conditions. Moreover, the universe is permeated by magnetic fields of varying strengths, ranging from $10^{-4}\,\text{Gauss}$ at the galactic center \cite{Crocker:2010xc} to $10^{16}\,\text{Gauss}$ on the surfaces of certain magnetars \cite{Olausen:2013bpa}. Massive compact objects (e.g., Sgr A$^\ast$ and M87) are often immersed in surrounding magnetic fields \cite{Olausen:2013bpa,Mori:2013yda,Kennea:2013dfa,2015llg..book..391M,Eatough:2013nva, 2018A&A...618L..10G,EventHorizonTelescope:2021srq,EventHorizonTelescope:2024hpu,EventHorizonTelescope:2024rju}. Thus, MPP is anticipated to be a ubiquitous astrophysical process.

Extensive studies of MPP have been conducted across several complex spacetime backgrounds, including Buchdahl stars \cite{Shaymatov:2024ibv}, magnetized Reissner-Nordström (RN) black holes \cite{Shaymatov:2022eyz}, magnetized Kerr black holes \cite{Chakraborty:2024aug}, Konoplya-Rezzolla-Zhidenko parametrized black holes \cite{Xamidov:2024wou}, higher-dimensional Myers-Perry black holes \cite{Shaymatov:2024fle}, multiple types of regular black holes \cite{Viththani:2024map}, and naked singularities \cite{Patel:2023efv}. In astrophysics, MPP can explain the origins and production mechanisms of extragalactic ultra-high-energy cosmic rays with energies exceeding $10^{20}\,\text{eV}$ \cite{Tursunov:2020juz,Tursunov:2022iqk}, which surpass the GZK cutoff \cite{Greisen:1966jv,Zatsepin:1966jv,PierreAuger:2017pzq,PierreAuger:2018qvk}. Additionally, MPP has been used to account for relativistic jets \cite{Stuchlik:2015nlt} and the radiative effects of gamma-ray burst jets \cite{1985ApJ...290...12W}.  These studies demonstrate that magnetic fields around black holes significantly enhance energy extraction efficiency, rendering MPP a highly promising mechanism for understanding high-energy phenomena near compact astrophysical objects \cite{Shaymatov:2022eyz,Shaymatov:2024fle,Shaymatov:2024ibv}. Related work and mechanisms include Ruffini and Wilson's work on energy extraction from rotating black holes via charge separation in magnetized plasmas (based on Kerr accretion) \cite{Ruffini:1975ne}, collisional Penrose process \cite{piran1975high}, superradiant scattering \cite{Teukolsky:1974yv}, the Blandford-Znajek (BZ) mechanism \cite{Blandford:1977ds}, magnetohydrodynamic (MHD) Penrose process \cite{1990ApJ...363..206T}, the Ba\~{n}ados-Silk-West (BSW) mechanism \cite{Banados:2009pr}, and the Comisso-Asenjo mechanism \cite{Comisso:2020ykg}.

Inspired by these works, in this paper, we explore a non-Kerr rotating spacetime, the Quevedo-Mashhoon (QM) spacetime, and perform an in-depth analysis of MPP within it. First introduced in 1991, the QM spacetime effectively models the external gravitational field of arbitrarily rotating deformed bodies \cite{Quevedo:1989rfm,Quevedo:1991zz}. Characterized by an infinite set of mass and current multipole moments, this metric is a stationary, axisymmetric solution to the vacuum Einstein equations within the Weyl-Lewis-Papapetrou class \cite{Weyl:1917gp,Lewis:1932zz,Papapetrou:1966zz} and contains a naked singularity. In the absence of rotation, it reduces to the Erez-Rosen solution \cite{Carmeli:2001ay}; for small rotations and an anomalous quadrupole moment, it approximates the Hartle-Thorn metric \cite{Boshkayev:2012ej}. The geometric properties, geodesics, and accelerated motion in the QM spacetime have been extensively studied \cite{Bini:2009cg}. Analogous spacetimes include the Manko-Novikov spacetime \cite{Novikov:1992gow}. We anticipate that MPP could reveal unique physical phenomena distinguishing the QM spacetime from Kerr black holes, thereby offering reliable observational signatures. 

The structure of this paper is as follows: Sec. II introduces the QM spacetime and an external uniform magnetic field, with the magnetic field configuration described by the Wald solution. We conduct a thorough investigation into how the anomalous quadrupole moment influences the near-horizon geometry and ergoregion. Sec. III explores the effects of the anomalous quadrupole moment and magnetic field on the negative-energy regions of charged particles through in-depth analysis of their motion. Sec. IV then examines the scenario where particles in the negative-energy region split into two via the MPP, calculating in detail how the anomalous quadrupole moment and magnetic field specifically affect the energy extraction efficiency of this process. The final section (Sec. V) presents our conclusions and discussions.

Throughout this work, we utilize the units $c = G = 4\pi \epsilon_0 = 1$, where $c, G$, and $\epsilon_0$ are the speed of light in vacuum, the Newton gravitational constant, and the vacuum permittivity, respectively.

\section{Non-Kerr rotating spacetime with external magnetic field}

\subsection{Quevedo-Mashhoon spacetime}

The Quevedo-Mashhoon (QM) metric \cite{Quevedo:1989rfm,Quevedo:1991zz} is an exact, stationary, axisymmetric, and asymptotically flat solution to the vacuum Einstein equations, encompassing arbitrary mass multipole moments. Beyond the mass and spin parameters, its most general form contains an infinite number of additional free parameters, which characterize deviations from the multipole moments of the Kerr metric. In this work, we focus on a special subclass of the QM metric first identified in \cite{1985PhLA..109...13Q}, which includes only the additional quadrupole moment parameter. In prolate spheroidal coordinates $(t, x, y, \phi)$ with $x \geq 1$ and $-1 \leq y \leq 1$, the metric reads \cite{2009esef.book.....S,Bini:2009cg}
\begin{align}
	ds^2  = -f (dt - \omega d\phi)^2 + \frac{k^2 e^{2 \gamma }}{f} (x^2 - y^2) \left(\frac{dx^2}{x^2-1} + \frac{dy^2}{1-y^2}\right) + \frac{k^2}{f} (x^2-1) (1-y^2) d\phi ^2,
\end{align}
where
\begin{align}
	f=\frac{A}{C} e^{-2 {\cal Q} P_2 Q_2}, \quad \omega = 2 a - 2 k \frac{D}{A} e^{2 {\cal Q} P_2 Q_2}, \quad e^{2 \gamma} =  \frac{1}{4} \left(1 + \frac{M}{k}\right)^2 \frac{A}{x^2 - 1} e^{2 \hat{\gamma}},
\end{align} 
and 
\begin{align}
A &= a_+ a_- + b_+ b_-, \nonumber\\ 
C &= a_+^2 + b_+^2, \nonumber\\
D &= \alpha x (1 - y^2) \left(e^{2 {\cal Q} \delta_+} + e^{2 {\cal Q} \delta_-}\right) a_+ + y (x^2 - 1) \left[1 - \alpha^2 e^{2 {\cal Q} (\delta_+ + \delta_-)}\right] b_+, \nonumber\\
\hat{\gamma} &= \frac{1}{2} (1 + {\cal Q})^2 \ln \frac{x^2 - 1}{x^2 - y^2} + 2 {\cal Q} (1 - P_2) Q_1 + {\cal Q}^2 (1 - P_2) \bigg[ (1 + P_2) (Q_1^2 - Q_2^2) \nonumber\\
&\quad + \frac{1}{2} (x^2 - 1) (2 Q_2^2 - 3 x Q_1 Q_2 + 3 Q_0 Q_2 - Q_2') \bigg].
\end{align}
Here $P_m(y)$ and $Q_m(x)$ are the $m$-th order Legendre polynomials of the first and second kind, respectively. Furthermore,
\begin{align}
a_{\pm} &= x \left[1 - \alpha^2 e^{2 {\cal Q}(\delta_+ + \delta_-)}\right] \pm \left[1 + \alpha^2 e^{2 {\cal Q}(\delta_+ + \delta_-)}\right], \nonumber\\
b_{\pm} &= \alpha y \left(e^{2 {\cal Q}\delta_+} + e^{2 {\cal Q}\delta_-}\right) \mp \alpha \left(e^{2 {\cal Q}\delta_+} - e^{2 {\cal Q}\delta_-}\right), \nonumber\\
\delta_{\pm} &= \frac{1}{2} \ln \frac{(x \pm y)^2}{x^2 - 1} + \frac{3}{2} \left(1 - y^2 \mp xy\right) + \frac{3}{4} \left[x \left(1 - y^2\right) \mp y \left(x^2 - 1\right)\right] \ln \frac{x - 1}{x + 1},
\end{align}
and the constants $k$ and $\alpha$ are defined as
\begin{align}
    \alpha = \frac{M-k}{a},\quad k = \sqrt{M^2 -a^2}.
\end{align}
The metric is characterized by three parameters: the mass $M$, the spin $a$, and the anomalous quadrupole moment parameter ${\cal Q}$.  The mass quadrupole moment of this spacetime is given by \cite{Bini:2009cg, Frutos-Alfaro:2016arb}
\begin{align}
    {\cal M}_2 = - M a^2 + \frac{2}{15} {\cal Q} M^3 \left(1- \frac{a^2}{M^2}\right)^{3/2}. \label{QuadrupoleMoment}
\end{align}
The metric reduces exactly to the Kerr metric when ${\cal Q} =0$, to the Erez-Rosen solution \cite{Carmeli:2001ay} when $a=0$, and approximates the Hartle-Thorn metric \cite{Boshkayev:2012ej} when both $a$ and ${\cal Q}$ are small. ${\cal Q} >0$ and ${\cal Q}<0$ correspond to spacetimes that are more prolate and more oblate than the Kerr one, respectively. All higher moments are also differ from their Kerr counterparts. From the above equation, the influence of ${\cal Q}$ is more significant for smaller spin $a$. In the limit $a\rightarrow M$, the correction from ${\cal Q}$ vanishes, and the metric reduces to the extreme Kerr spacetime regardless of the value of ${\cal Q}$. 

It will be more convenient to work with the Boyer-Lindquist coordinates $(r, \theta )$, which are related to the prolate spherodial coordinates $(x, y)$ by the relations 
\begin{align}
	x = \frac{r - M}{k}, \quad y = \cos \theta,\
\end{align}
and then
\begin{align} \label{BLmetric}
	ds^2 = -f (dt - \omega d \phi )^2 + \frac{e^{2 \gamma } \rho^2}{f \Delta } dr^2 + \frac{e^{2 \gamma}  \rho^2 }{f} d \theta^2 + \frac{\Delta \sin^2 \theta }{f} d \phi^2,
\end{align}
where $\rho^2 = (r-M)^2 - k^2 \cos^2 \theta$ and $\Delta=(r-M)^2 -k^2$. The (would-be) event horizon lies at $r = r_h=M+k$. The ergoregion is defined as the region outside the horizon where $g_{tt}>0$, from which energy can be extracted via the Penrose process.

Due to its deviations from the Kerr metric, the QM spacetime exhibits certain pathologies as implied by the no-hair theorem \cite{Carter:1971zc,Robinson:1975bv,Chrusciel:2008js}. In fact, when ${\cal Q} \neq 0$, the event horizon is disrupted by a naked singularity. Consequently, the QM metric generally does not describe a black hole spacetime in a standard sense. Nevertheless, it is appropriate to use the QM metric to describe the exterior gravitational field of a rotating body with an arbitrary quadrupole moment, where the naked singularity is obscured by matter or quantum effects and thus does not persist. 

\begin{figure}[htb]
	\centering
        \includegraphics[width=0.31\linewidth]{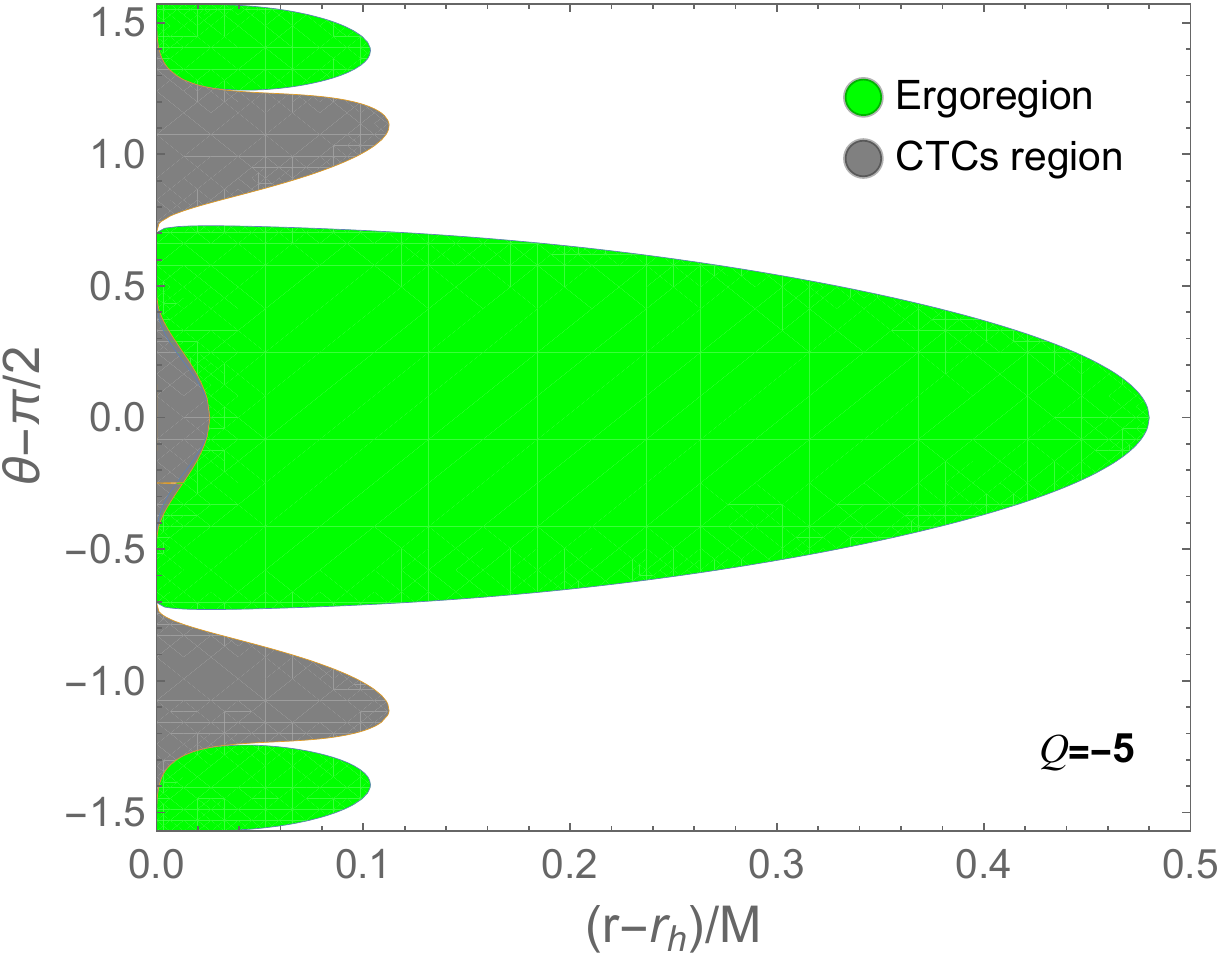}\quad
        \includegraphics[width=0.31\linewidth]{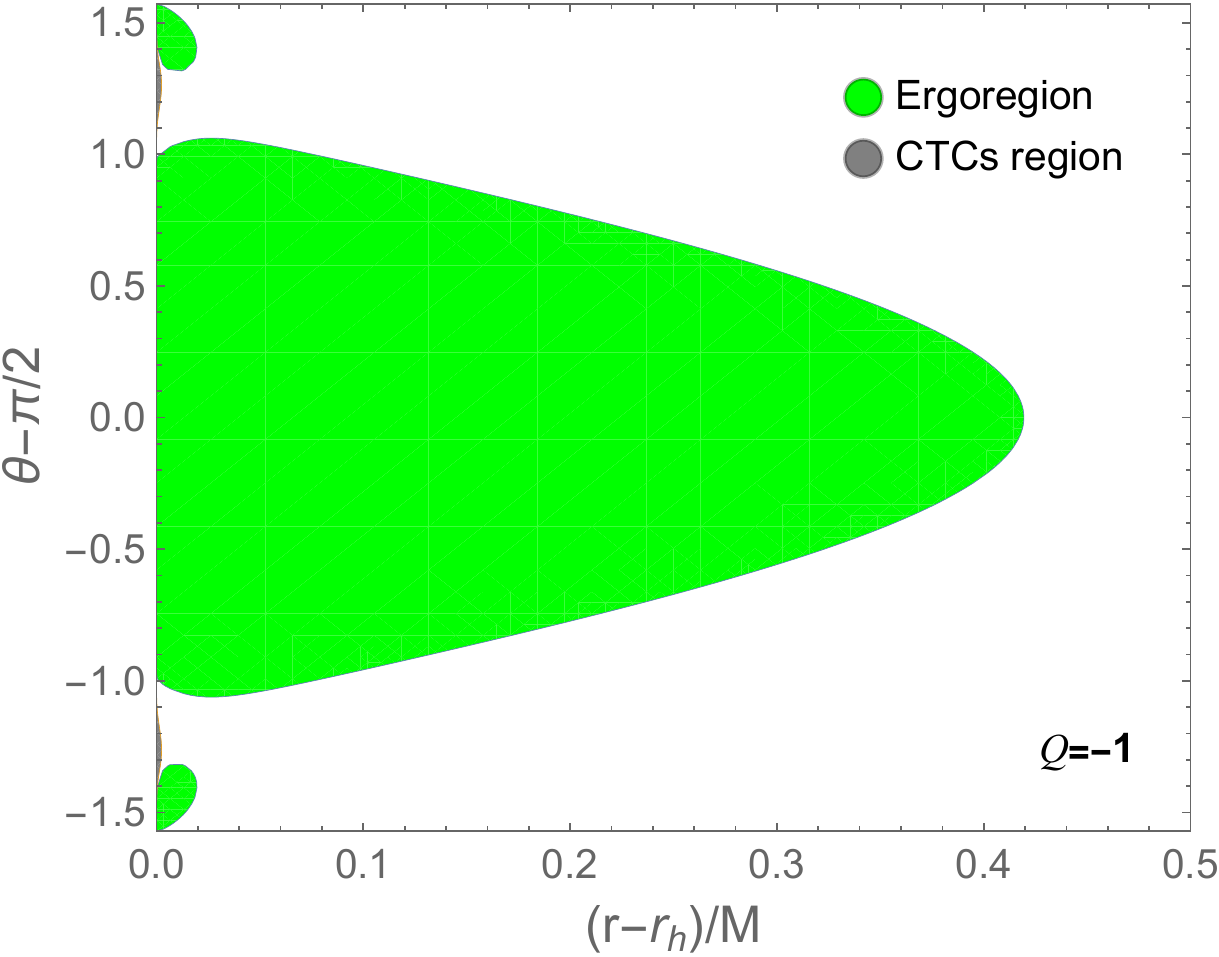}\quad
	\includegraphics[width=0.31\linewidth]{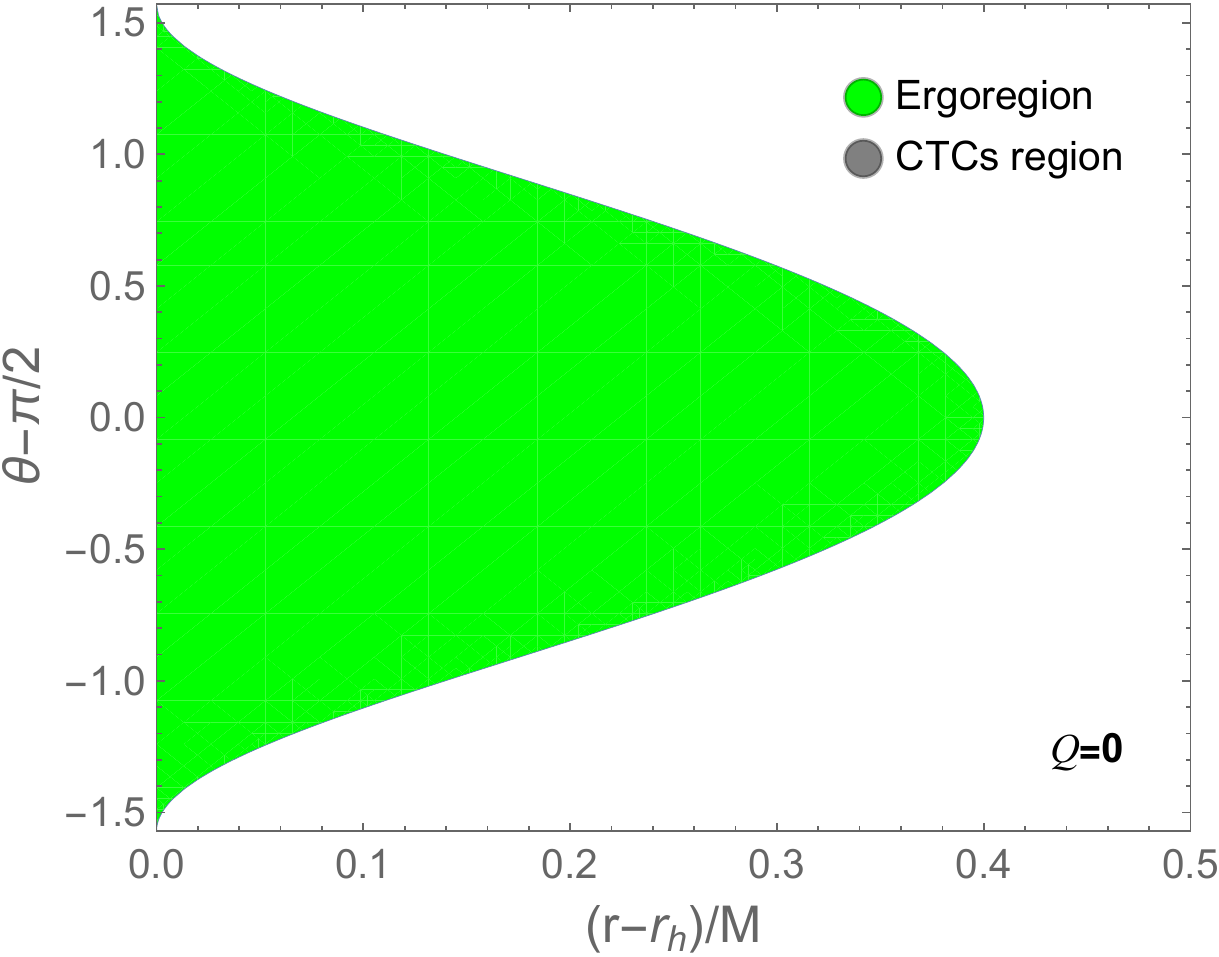}\\
 \vspace{0.5cm}
        \includegraphics[width=0.31\linewidth]{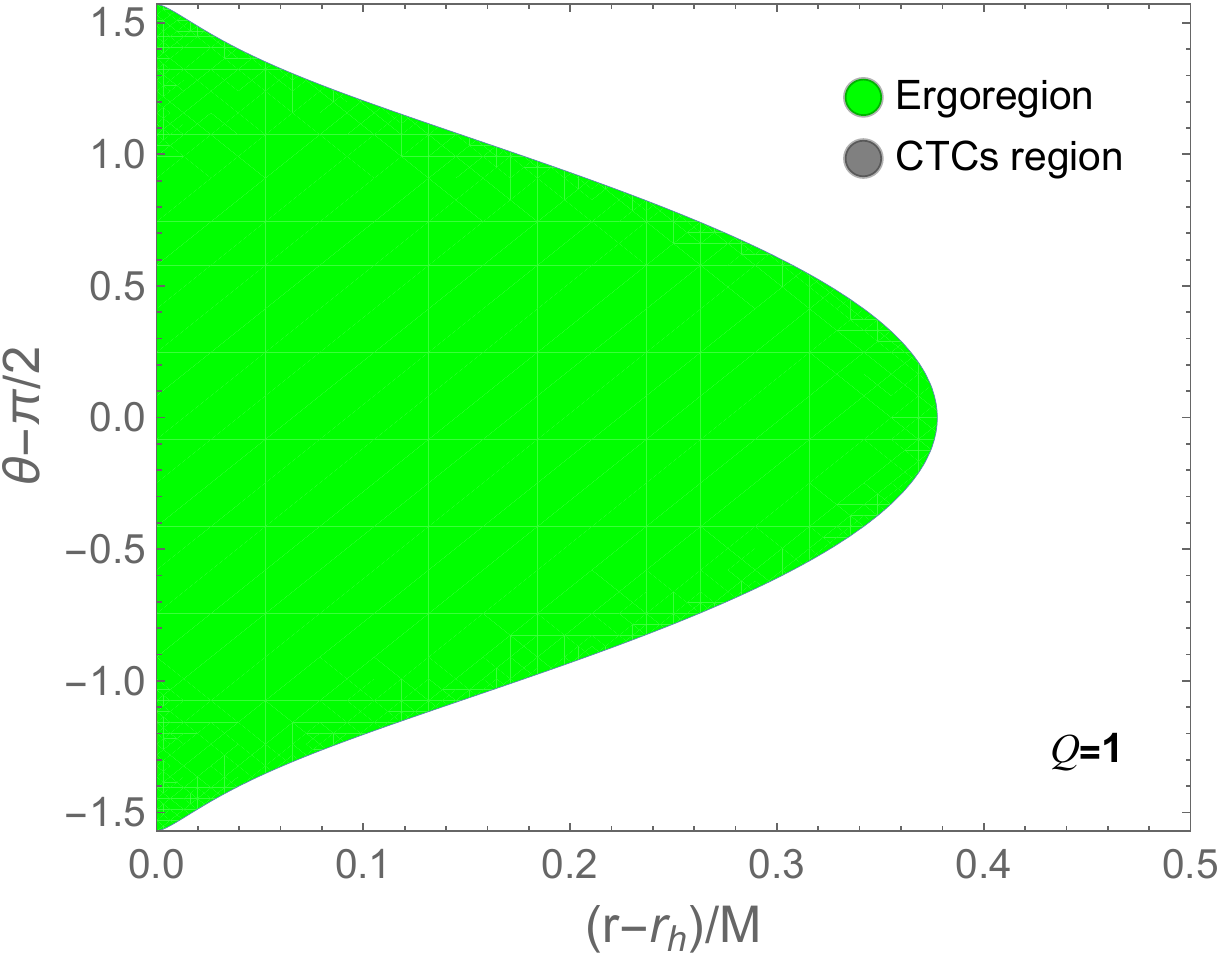}\quad
        \includegraphics[width=0.31\linewidth]{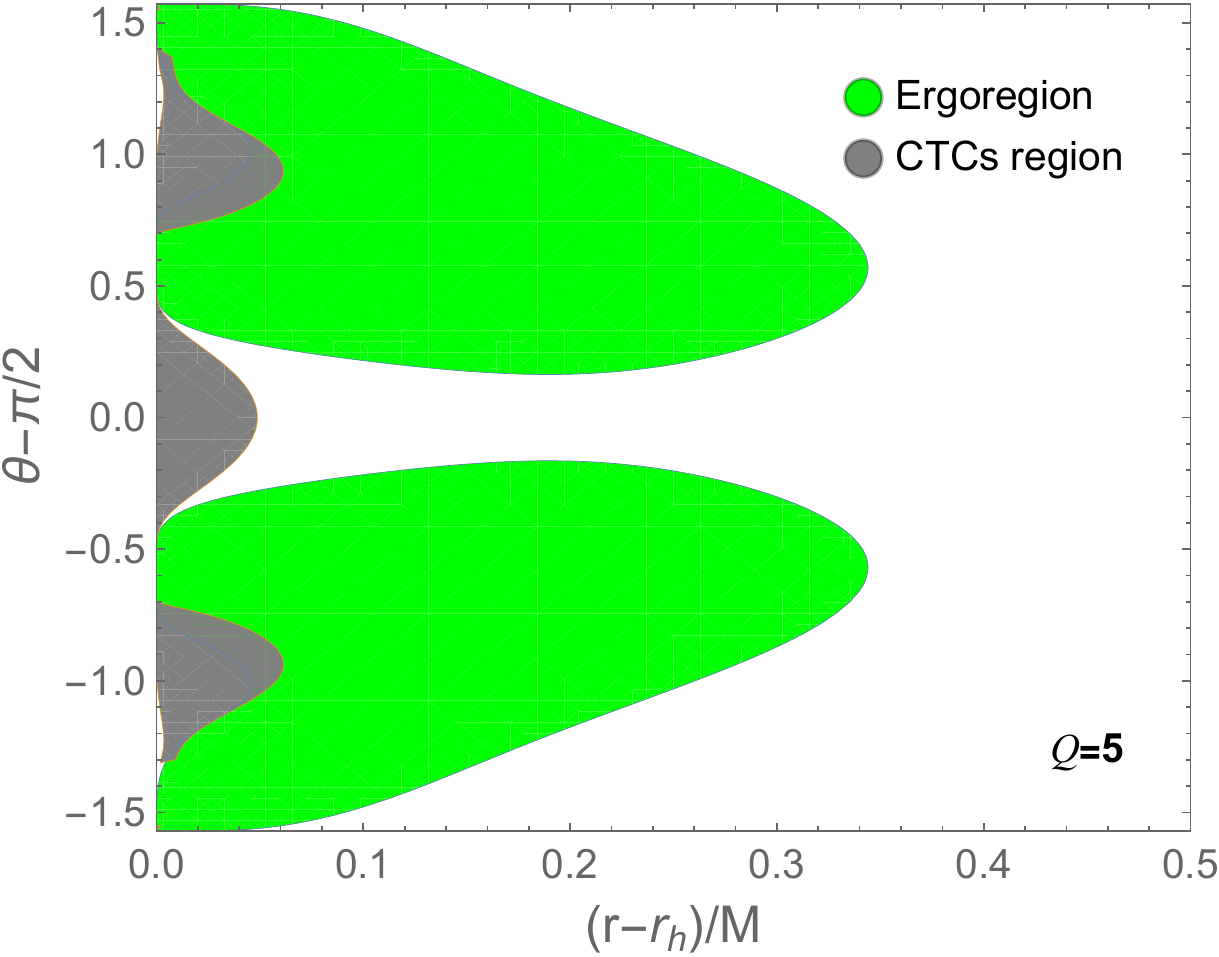}
    \caption{Shapes of the ergoregion and closed time-like curves (CTCs) region for different ${\cal Q}$. We set $a = 0.8$. For the Kerr case (${\cal Q}=0$) and small $|{\cal Q}|$, there is no CTCs region. }
	\label{ErgoregionCTCs}
\end{figure}

Beyond the possible naked singularity, there may also be regions outside the would-be horizon where $g_{\phi\phi}<0$ such that closed time-like curves (CTCs) can in principle exist. The influence of ${\cal Q}$ on the ergoregion and the CTCs region is shown in Fig. \ref{ErgoregionCTCs}. When $|{\cal Q}|$ exceeds a certain critical value, CTCs region emerges adjacent to the horizon, and both the CTCs region and the ergoregion exhibit a multi-lobe structure. More precisely, for sufficiently large ${\cal Q}>0$, the ergoregion splits into two separate parts and does not intersect the equatorial plane $\theta = \pi /2$, while the CTCs region splits into three separate parts. For sufficiently large ${\cal Q}<0$, both kinds of regions split into three separate parts, and the ergoregion always intersects the equatorial plane. Although theoretically allowed, the CTCs region is generally regarded as unphysical. A spacetime without CTCs region can be constructed by adding an inner boundary in the spacetime and using only the portion of the QM solution exterior to that boundary.

\subsection{External uniform magnetic field}

We consider the body to be immersed in an external uniform magnetic field. In our units, one can define a characteristic magnetic field strength $B_M \equiv 1/M$, at which order the electromagnetic energy is comparable to the gravitational energy of the body \cite{Aliev:1989wz,Melvin:1963qx}. Restoring physical units, we have
\begin{align}
    B_M \sim 2.35 \times 10^{19} \left(\frac{M_\odot}{M}\right) \text{Gauss}, \label{BM}
\end{align}
with $M_\odot$ being the solar mass. From this expression, $B_M$ is inversely proportional to the body's mass and generally takes an extremely high value. In our universe, magnetic fields around black holes or other compact objects are typically many orders of magnitude weaker than $B_M$, so generally we have $B M \ll 1$ and their backreaction on the spacetime can be neglected. However, the configuration of the external magnetic field is distorted by the compact object. In Maxwell's theory, the vacuum Maxwell equations in the Lorentz gauge admit Killing vectors as solutions. Following Wald's solution \cite{Wald:1974np}, we consider the vector potential to take the following form
\begin{align}
    A = a B \partial_t + \frac{B}{2} \partial_\phi, \label{ElectromagneticPotential}
\end{align}
which describes a sourceless (no electric or magnetic charge) and asymptotically uniform magnetic field of strength $|B|$, aligned ($B>0$) or anti-aligned ($B<0$) with the spacetime's symmetry axis. Moreover, we adopt the gauge $A_t (r\rightarrow \infty)=0$ \cite{Aliev:2002nw}, such that the resulting vector potential is
\begin{align}
    A_t &= a B (1 + g_{tt})  + \frac{B}{2} g_{t\phi},\nonumber\\
    A_\phi &= a B g_{t\phi} + \frac{B}{2} g_{\phi\phi},\nonumber\\
    A_r &= A_\theta =0. \label{ElectromagneticPotential-1}
\end{align}

Although any astrophysical magnetic field is intrinsically too weak to alter the background spacetime metric, its influence on the motion of charged particles cannot be neglected. In particular, the negative-energy region is extended beyond the ergoregion, as shown below.

\section{Motion of charged particles and negative energy regions}

\subsection{Equations of motion}

The motion of a charged particle in the spacetime can be described by the following Lagrangian
\begin{align}
    {\cal L} = \frac{1}{2} m g_{\mu \nu} \dot{x}^\mu \dot{x}^\nu + q A_\mu \dot{x}^\mu, \label{Lagrangian}
\end{align}
where parameters $m$ and $q$ are the mass and electric charge of the particle, respectively. The dot denotes the derivative with respect to the proper time $\tau$, i.e., $\dot{x}^\mu \equiv d x^\mu /d\tau$.

Due to the axial-symmetry of the spacetime, the Lagrangian (\ref{Lagrangian}) does not depend on two coordinates $\{t, \phi\}$ explicitly, so their conjugate momenta are constants of motion, that is
\begin{align}
    \frac{\partial {\cal L}}{\partial \dot{t}} & = m U_t + q A_t \equiv - E, \label{Energy}\\
    \frac{\partial {\cal L}}{\partial \dot{\phi}} &= m U_\phi + q A_\phi \equiv L, \label{AngularMomentum}
\end{align}
where $U^\mu \equiv \dot{x}^\mu$ is the 4-velocity of the particle. The two constants $E$ and $L$ are, respectively, the energy and the $z$-component of the angular momentum of the particle as seen from observers at infinity.

For a particle moving in the equatorial plane, its 4-velocity can be expressed as $U^\mu = \dot{t} (1, v, 0, \Omega)$, where $v \equiv dr/dt$ and $\Omega \equiv d\phi/dt$ are the radial and angular velocities measured at infinity, respectively. With this notation, the renormalization condition of the 4-velocity, i.e., $U_\mu U^\mu = -1$, can be written as
\begin{align}
   (\dot{t} )^2 \left(g_{tt}  + g_{rr} v^2  + g_{\phi\phi} \Omega^2 + 2 g_{t\phi} \Omega\right) = -1,\label{Normalization}
\end{align}
from which $\Omega$ can be expressed as
\begin{align}
    \Omega = -\frac{g_{t\phi}}{g_{\phi\phi}} \pm \sqrt{\left(\frac{g_{t\phi}}{g_{\phi\phi}}\right)^2 - \frac{1}{g_{\phi\phi}} \left[g_{tt}  + g_{rr} v^2 + \frac{1}{(\dot{t})^2}\right]}, \label{AngularVelocity}
\end{align}
where the sign ``$\pm$" stands for co-rotating and counter-rotating cases, respectively. From this expression, for any physical particle moving in the equatorial plane, its possible angular velocity $\Omega$ is limited to be
\begin{align}
    \Omega_- \leq \Omega \leq \Omega_+, \qquad \Omega_\pm = -\frac{g_{t\phi}}{g_{\phi\phi}} \pm \sqrt{\left(\frac{g_{t\phi}}{g_{\phi\phi}}\right)^2 - \frac{1}{g_{\phi\phi}} \left(g_{tt}  + g_{rr} v^2 \right)},
\end{align}
with $\Omega_\pm$ corresponding to the photon motion, i.e., $U_\mu U^\mu =0$.

\subsection{Effective potential}

With Eqs. (\ref{Energy})(\ref{AngularMomentum}),  the normalization condition (\ref{Normalization}) can be written as a quadratic equation of the specific energy $e \equiv E/m$,
\begin{align}
    e^2 - 2 \beta e + \gamma=0,
\end{align}
where 
\begin{align}
    \beta &= - \bar{q} A_t - \frac{g_{t\phi}}{g_{\phi\phi}} (\ell - \bar{q} A_\phi),\\
    \gamma &= \bar{q} A_t \left[\bar{q} A_t +2 \frac{g_{t\phi}}{g_{\phi\phi}} (\ell - \bar{q} A_\phi)\right] + \frac{g_{tt}}{g_{\phi\phi}} (\ell - \bar{q} A_\phi)^2 + \frac{g_{rr} (\dot{r})^2 + 1}{g^{tt}},
\end{align}
and $\ell\equiv L/m$ and $\bar{q} \equiv q/m$ are the specific angular momentum and the charge-to-mass-ratio of the particle, respectively. Solving this equation, we have 
\begin{align}
    e =  - \bar{q} A_t - \frac{g_{t\phi}}{g_{\phi\phi}} (\ell - \bar{q} A_\phi) + \sqrt{-\frac{1}{g^{tt}} \left[\frac{(\ell - \bar{q} A_\phi)^2}{g_{\phi\phi}} + g_{rr} \dot{r}^2  + 1 \right]}. \label{Energy1}
\end{align}
Here we have chosen the positive sign for the radical for the following reason:  A zero angular momentum observer (ZAMO)  is an interesting special type of observer that can live anywhere outside the horizon, including within the ergoregion. The 4-velocity of a ZAMO is $u_Z = u^t_Z \left(\partial_t - \frac{g_{t\phi}}{g_{\phi\phi}} \partial_\phi\right)$. Although a particle may have negative energy measured by observers at infinity ($e<0$), it must always possess positive energy as measured by a local ZAMO, i.e., $- p_\mu u_Z^\mu =  m u^t_Z \left[e + \bar{q} A_t + \frac{g_{t\phi}}{g_{\phi\phi}} (\ell - \bar{q} A_\phi)\right] >0 $. This yields the constraint $e > - \bar{q} A_t - \frac{g_{t\phi}}{g_{\phi\phi}} (\ell - \bar{q} A_\phi)$. For further discussions on this point, refer to \cite{Gupta:2021vww}.

With (\ref{Energy1}), one can define an effective potential as
\begin{align}
    V_{\rm eff} (r) &\equiv e (\dot{r}= \dot{\theta} = 0) \nonumber\\
    &= - \bar{q} A_t - \frac{g_{t\phi}}{g_{\phi\phi}} (\ell - \bar{q} A_\phi) + \sqrt{-\frac{1}{g^{tt}} \left[\frac{(\ell - \bar{q} A_\phi)^2}{g_{\phi\phi}} + 1 \right]},\label{EffectivePotential}
\end{align}
which gives the minimum specific energy that the particle must possess: a particle with specific energy $e$ can only move in regions where $e \geq V_{\rm eff}(r)$. The fact that $V_{\rm eff} (r)$ may be negative in certain regions outside the horizon implies that $e$ can be negative there, enabling energy extraction via the Penrose process. In the absence of the electromagnetic field $A_t = A_\phi =0$, $e<0$ is only possible within the ergoregion, which is where the purely mechanical Penrose process occurs. When the electromagnetic interaction is incorporated, the negative-energy region (NER), defined by the condition $V_{\rm eff} (r) < 0$, may extend beyond the ergoregion. 

\subsection{Negative-energy region}

As discussed above, the negative-energy region (NER) is defined by the condition $V_{\rm eff} (r) < 0$, with its boundary (referred to as the zero-energy surface) determined by the equation $V_{\rm eff} (r) =0$. From Eq. (\ref{EffectivePotential}), $V_{\rm eff} (r)$ depends on a set of parameters, $\{M, a, {\cal Q}, B, \bar{q}, \ell \} $. Additonally, $V_{\rm eff} (r)$ depends on $B$ and $\bar{q}$ through their product $\bar{q} B$.

For convenience, we set the body's mass $M=1$ in what follows, such that  all quantities discussed below are measured in units of $M$. 

\begin{figure}[htb]
	\centering
        \includegraphics[width=0.31\linewidth]{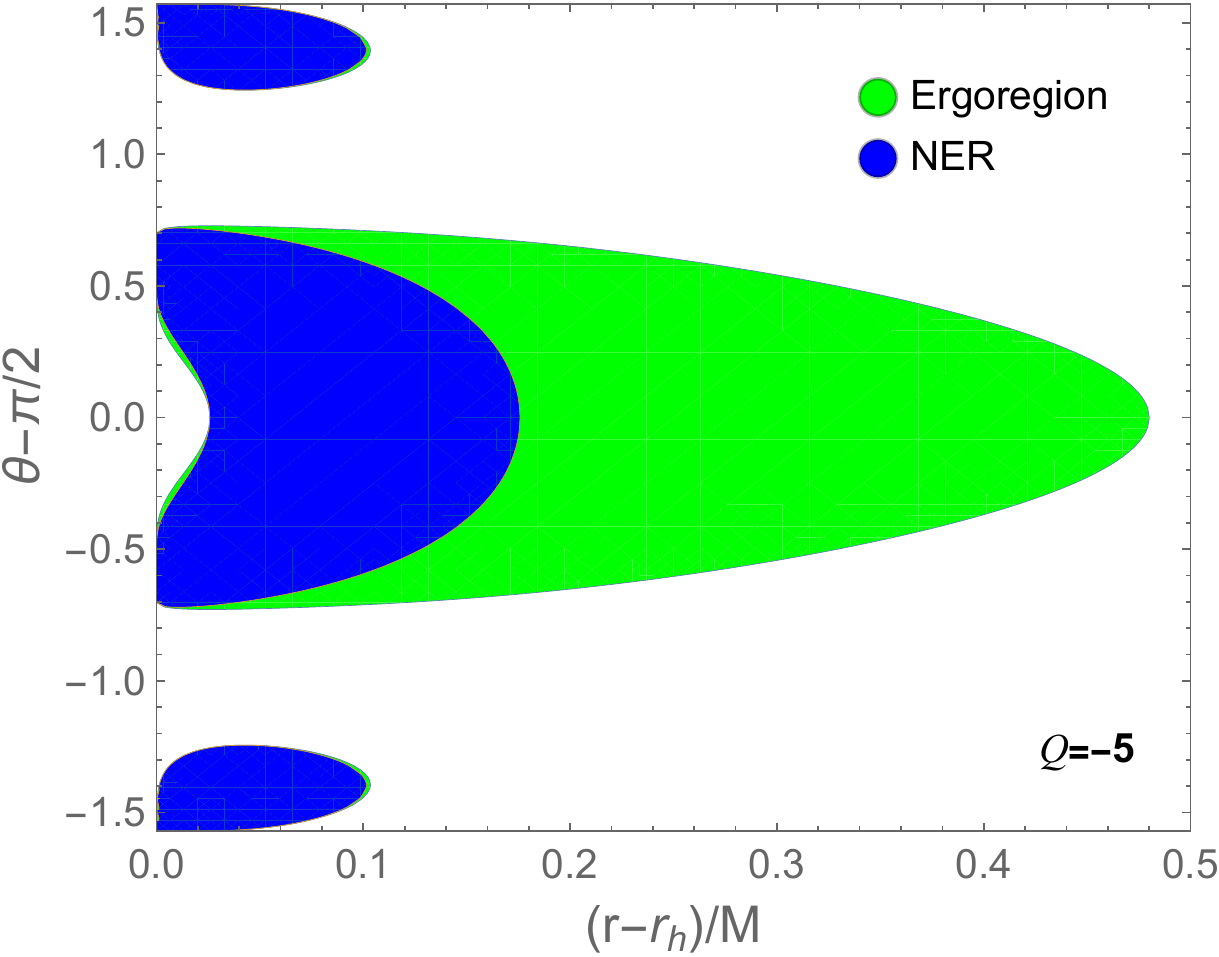}\quad
        \includegraphics[width=0.31\linewidth]{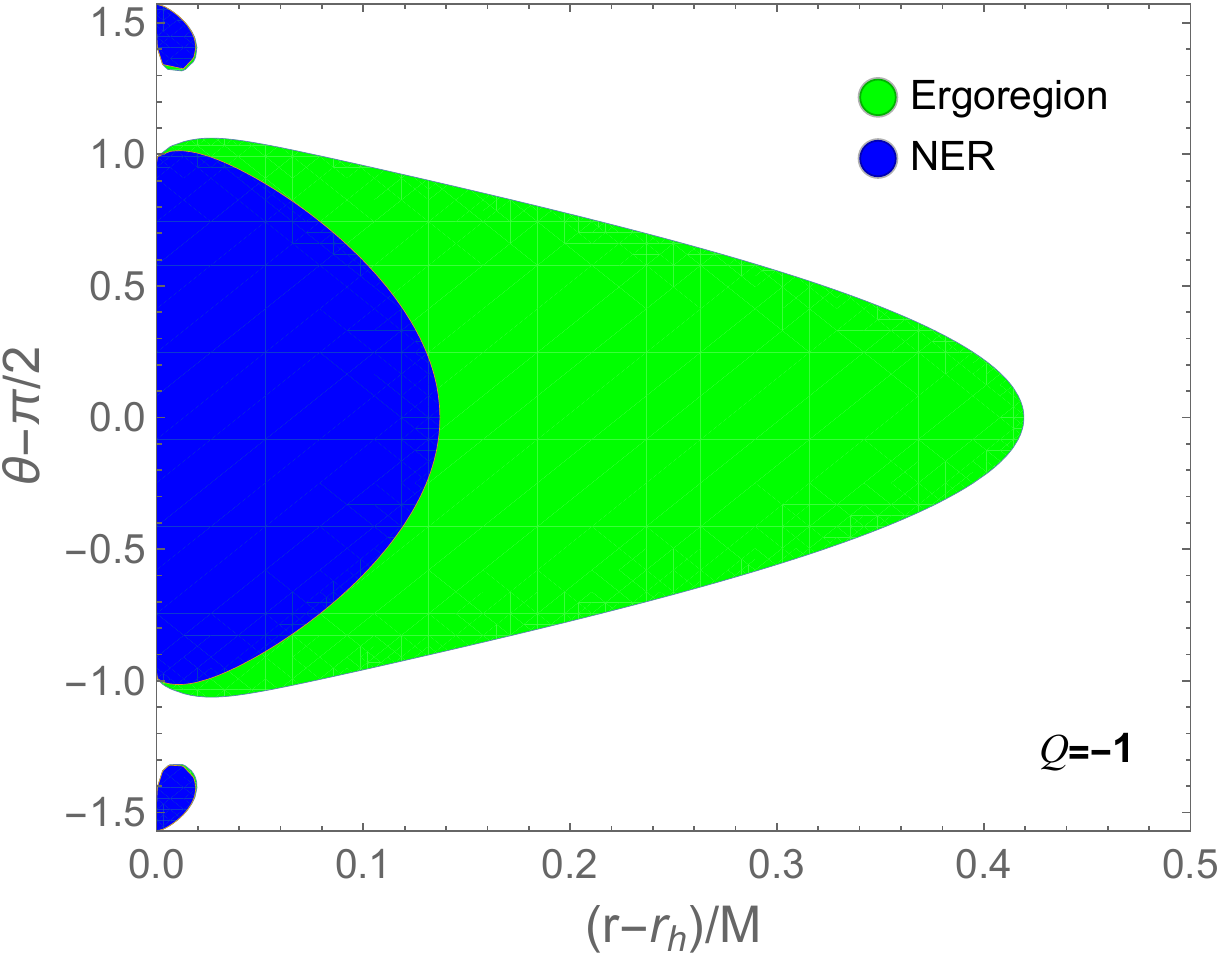}\quad
	\includegraphics[width=0.31\linewidth]{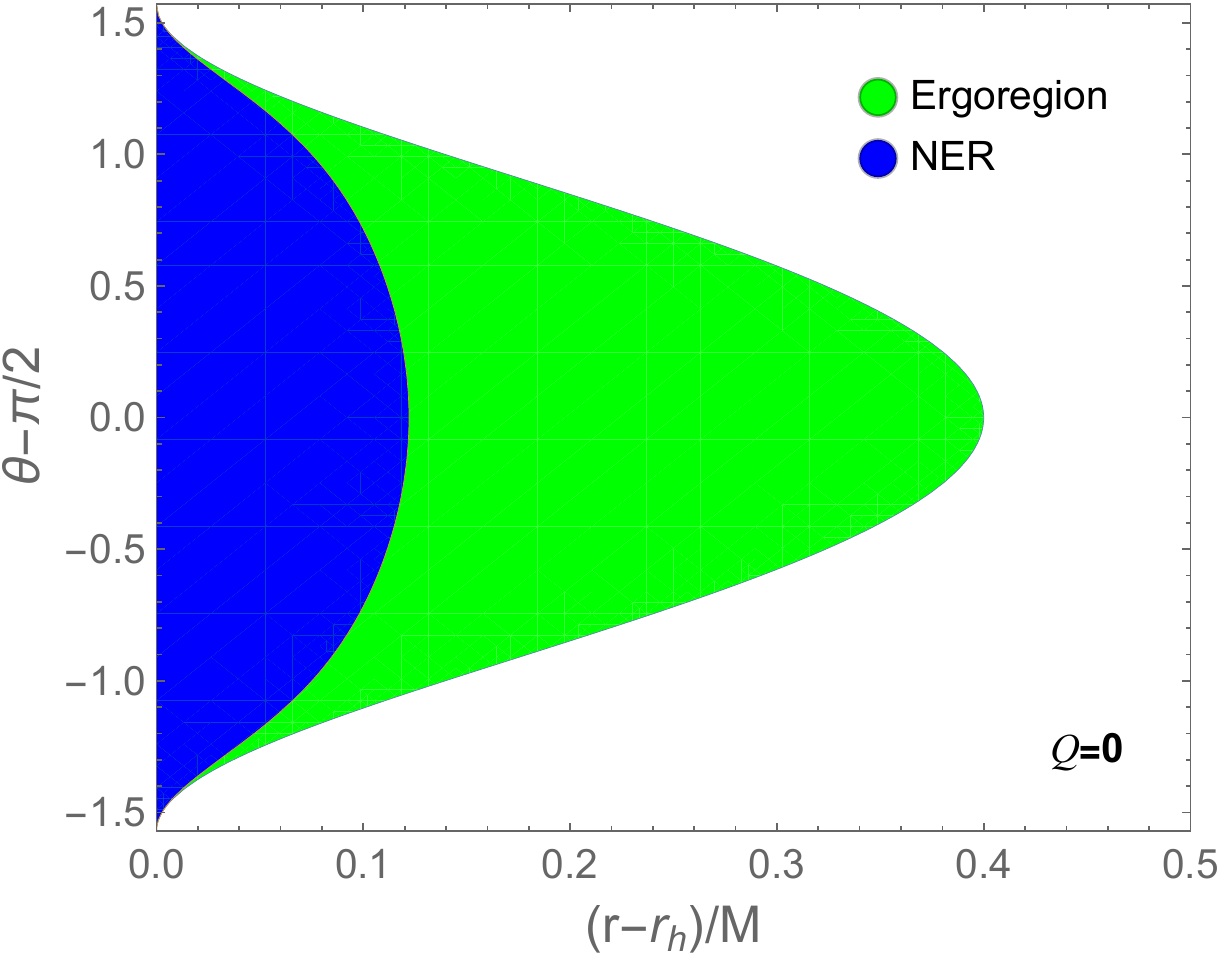}\\
 \vspace{0.5cm}
        \includegraphics[width=0.31\linewidth]{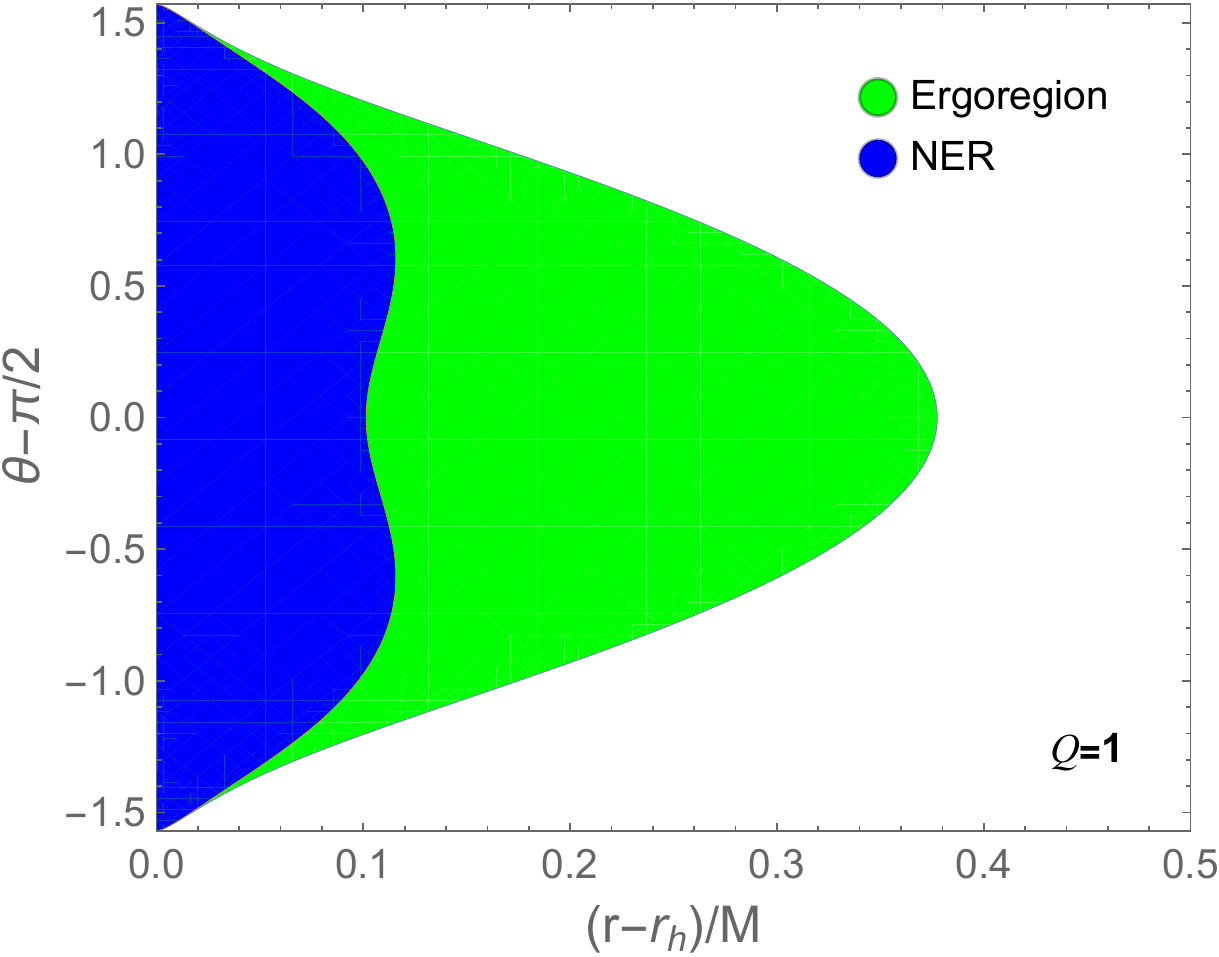}\quad
        \includegraphics[width=0.31\linewidth]{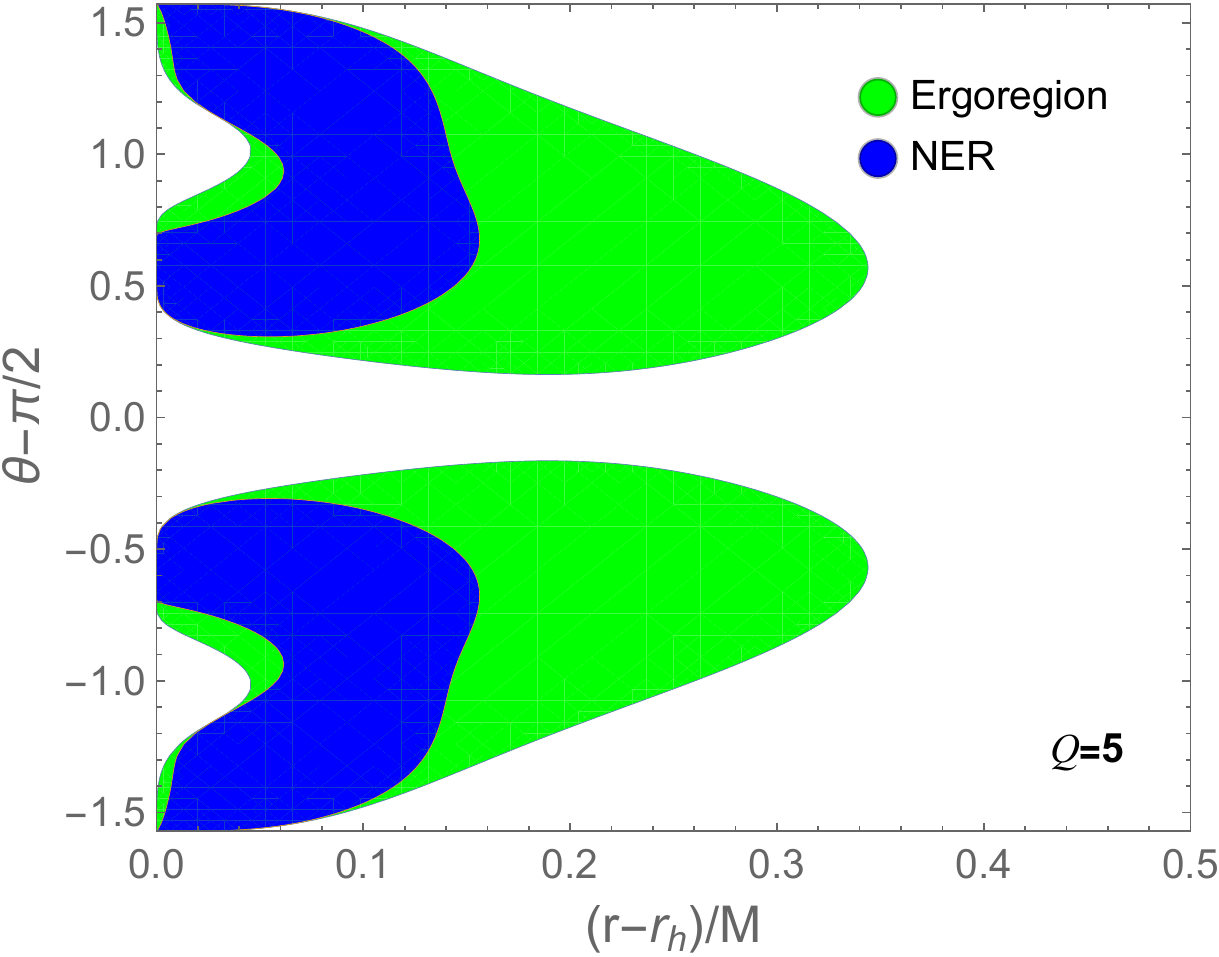}\quad
        \caption{Shapes of the negative-energy region (NER) and ergoregion for different ${\cal Q}$ with $\bar{q} B=0$. We set $a = 0.8$ and $\ell=-1$. In plotting the NER, the CTCs region has been excluded.}
	\label{NERQ-1}
\end{figure}

\begin{figure}[htb]
	\centering
        \includegraphics[width=0.31\linewidth]{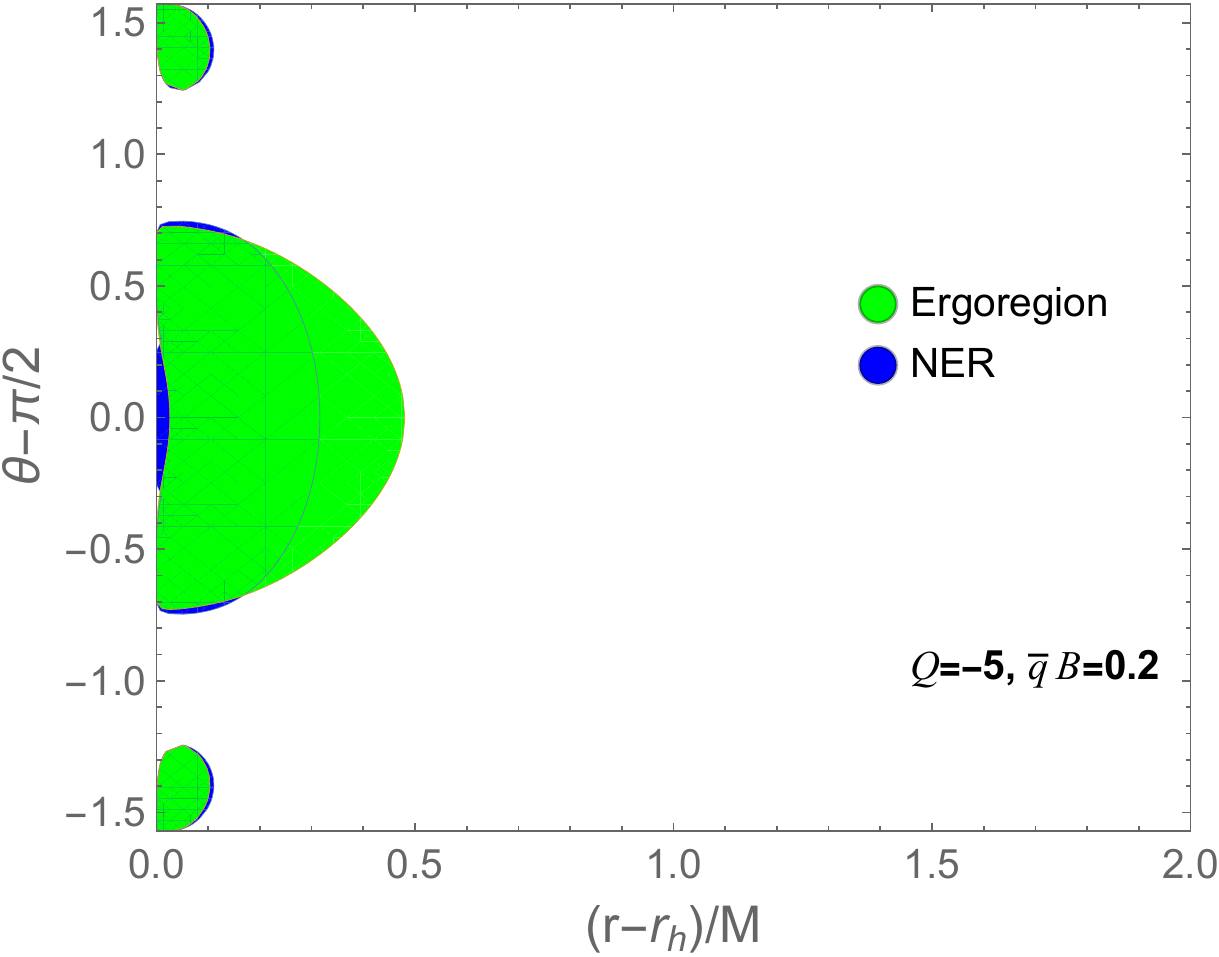}\quad
        \includegraphics[width=0.31\linewidth]{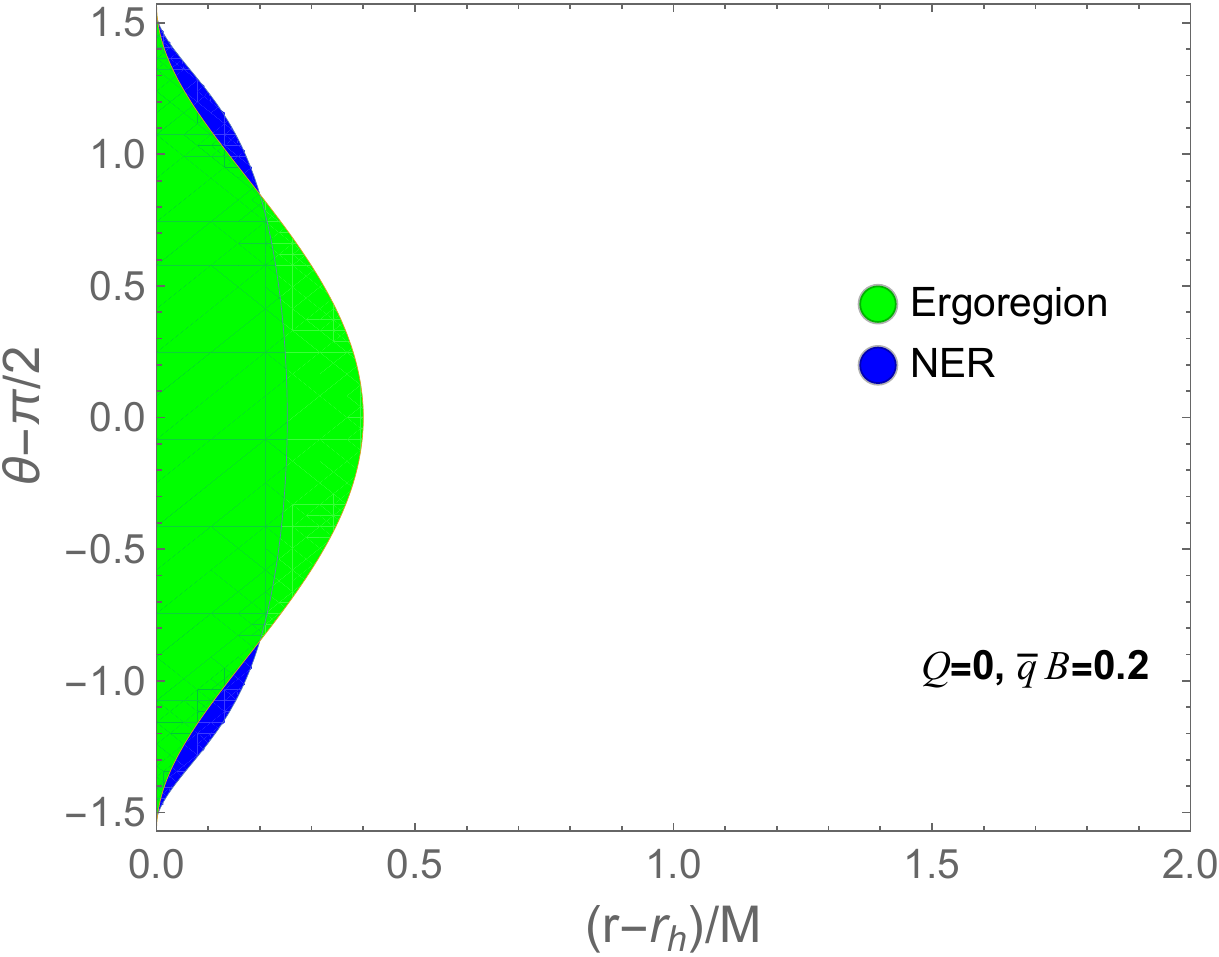}\quad
        \includegraphics[width=0.31\linewidth]{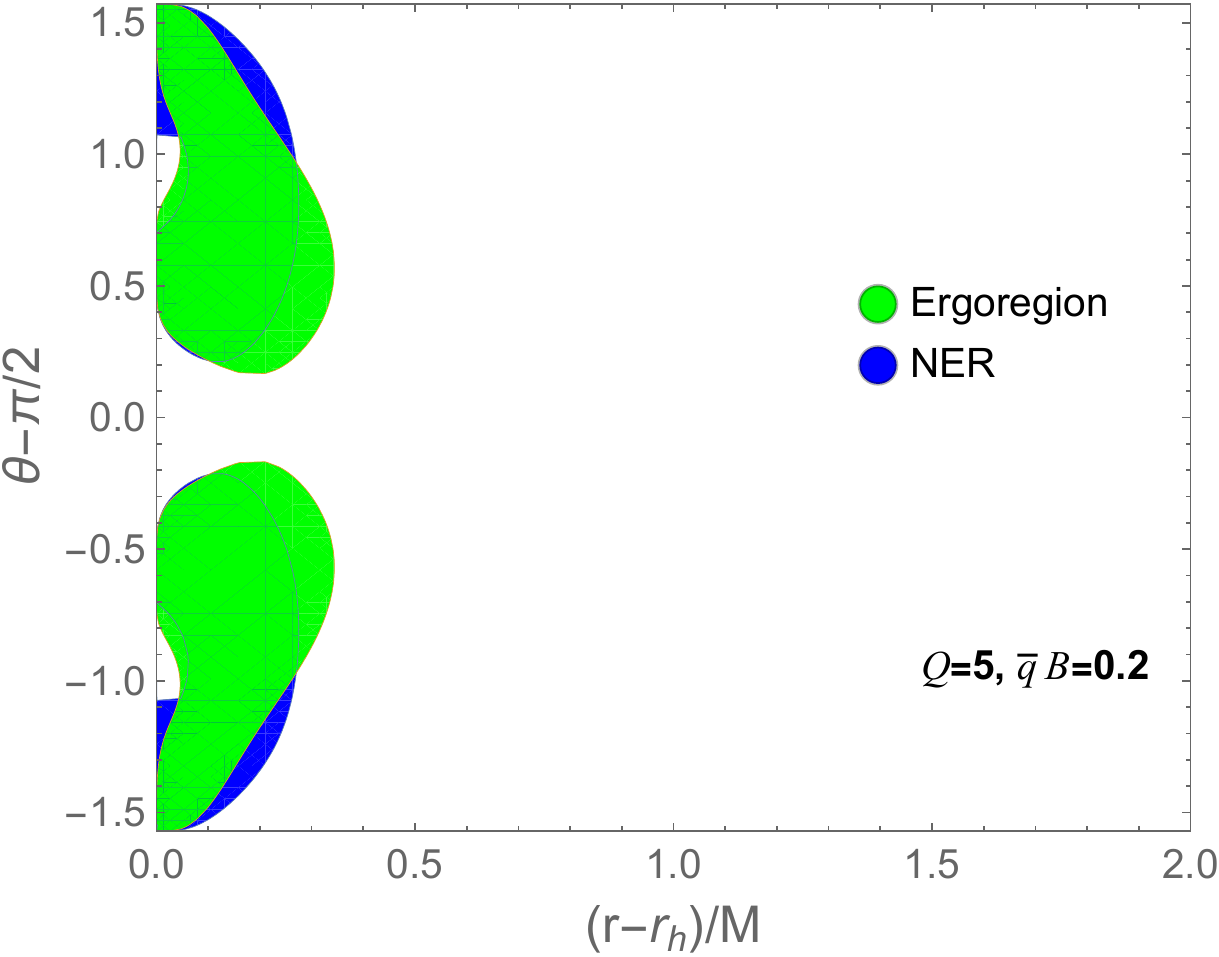}\\
       \vspace{0.5cm}
       
        \includegraphics[width=0.31\linewidth]{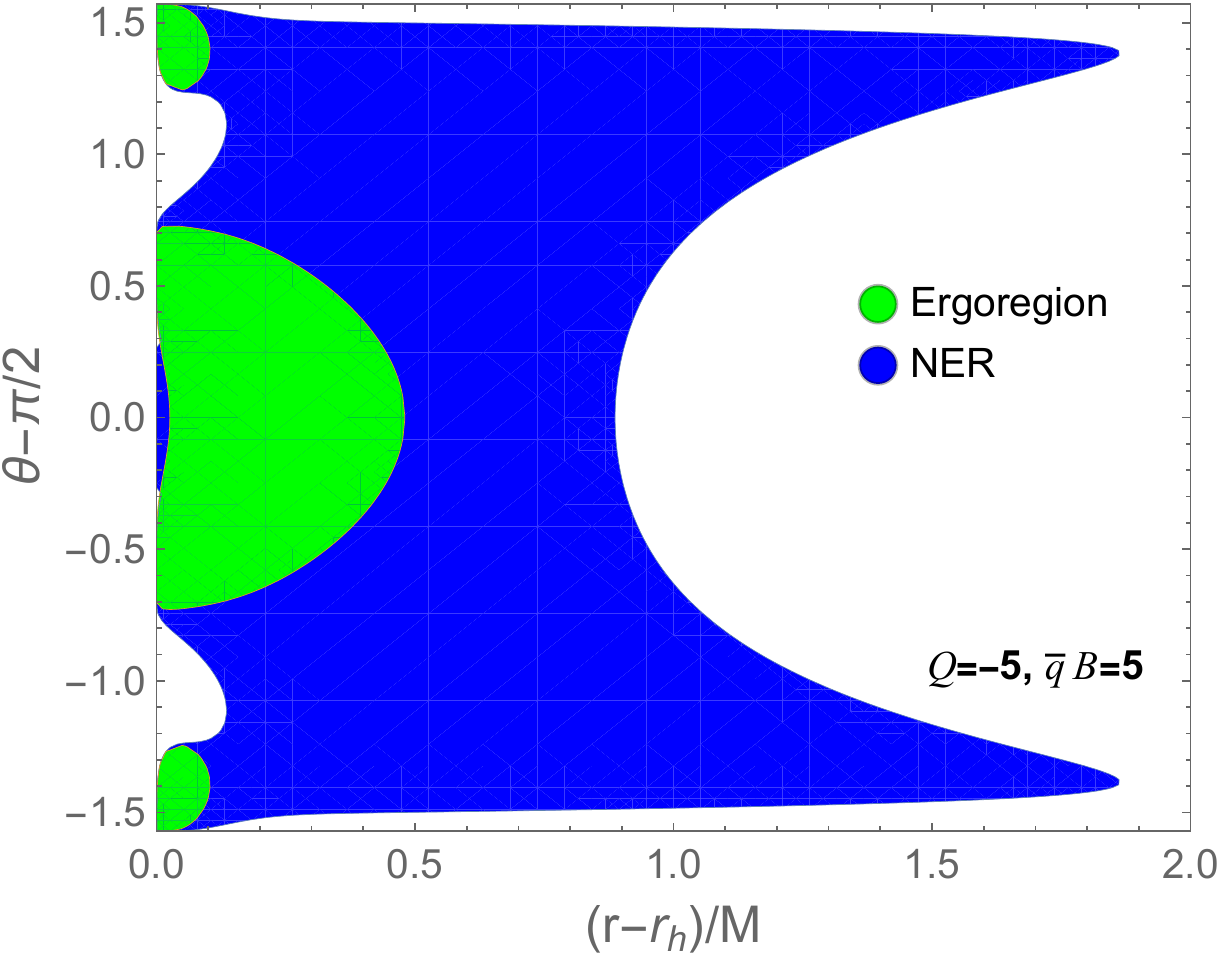}\quad
        \includegraphics[width=0.31\linewidth]{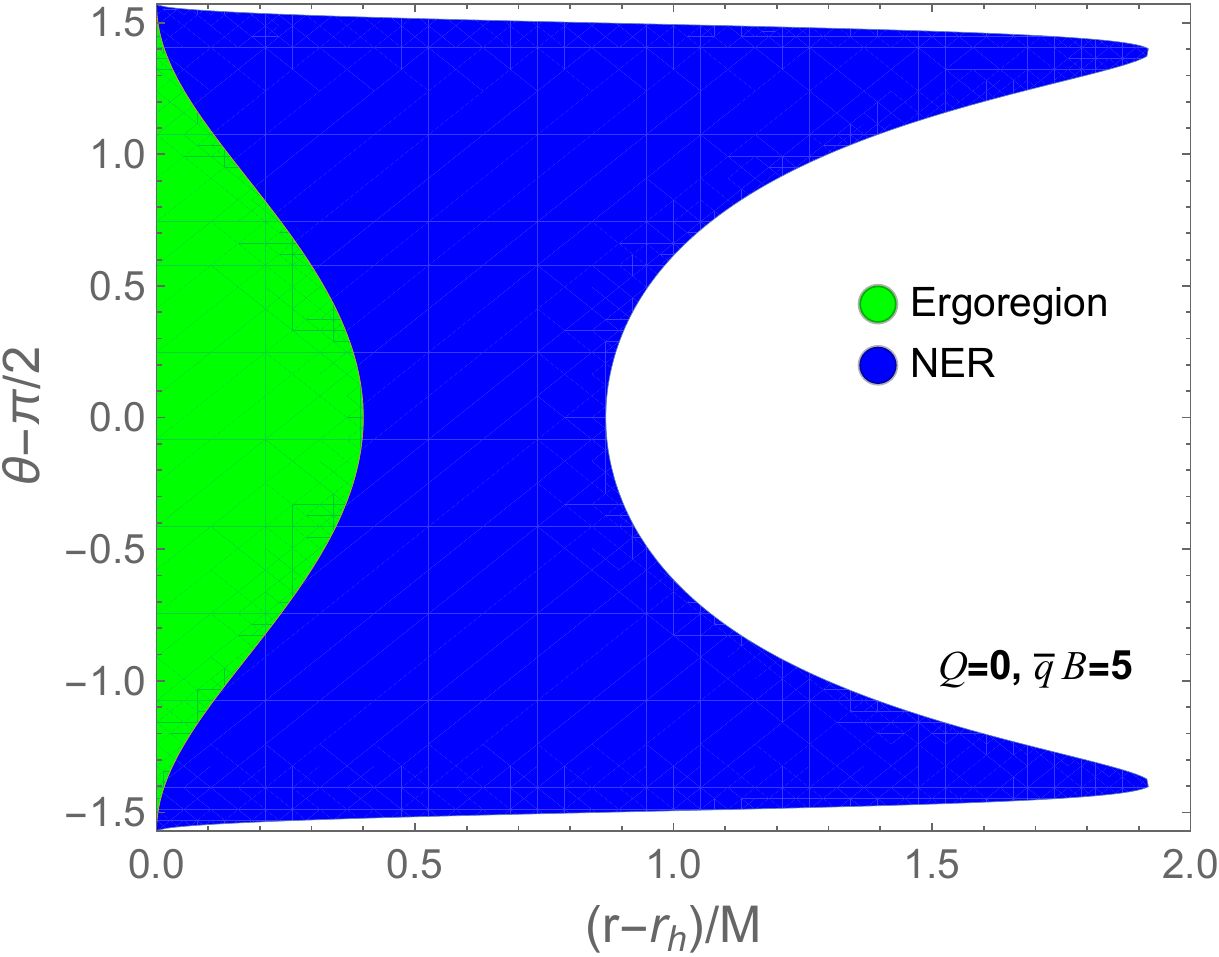}\quad
        \includegraphics[width=0.31\linewidth]{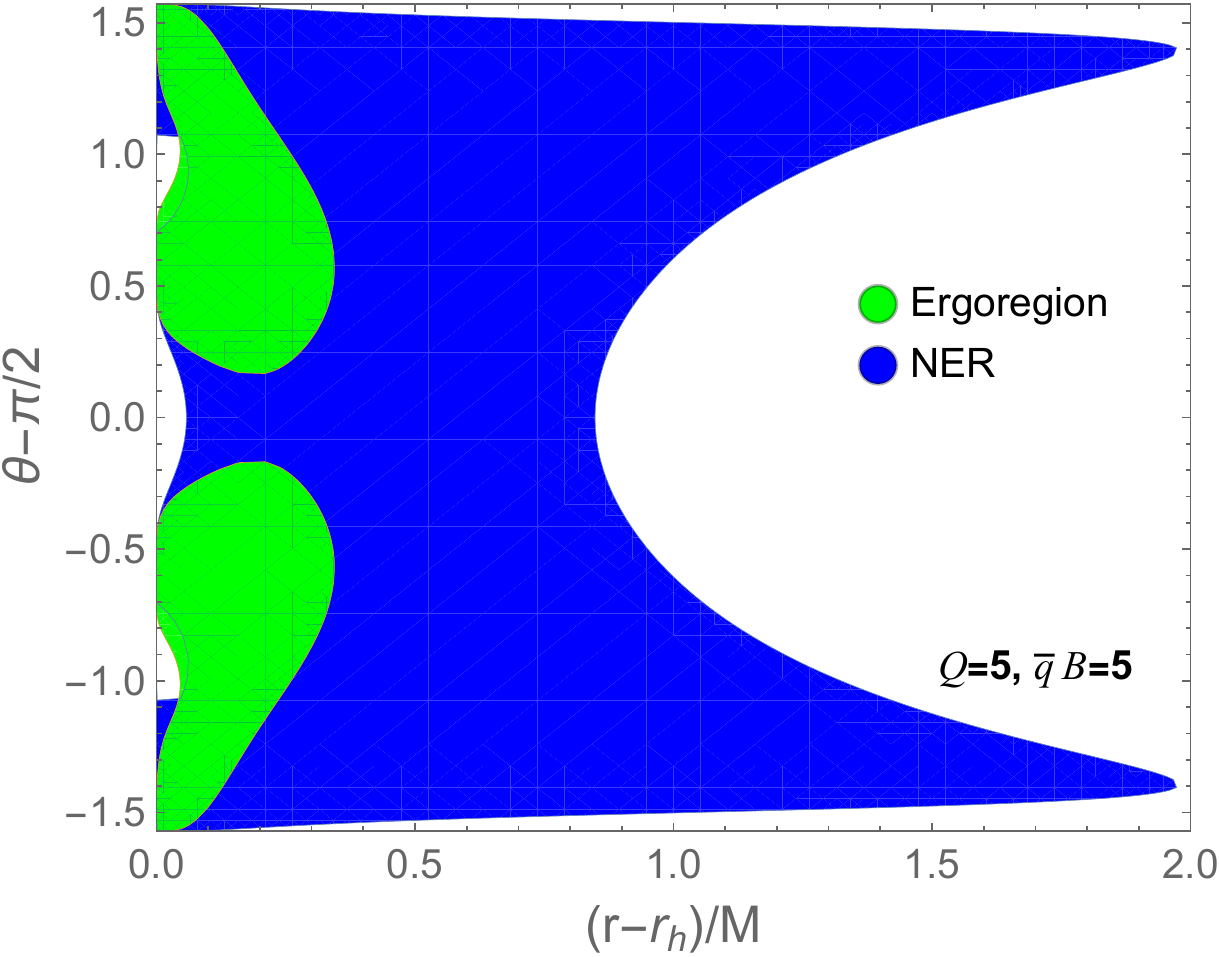}\\
        \vspace{0.5cm}
        \includegraphics[width=0.31\linewidth]{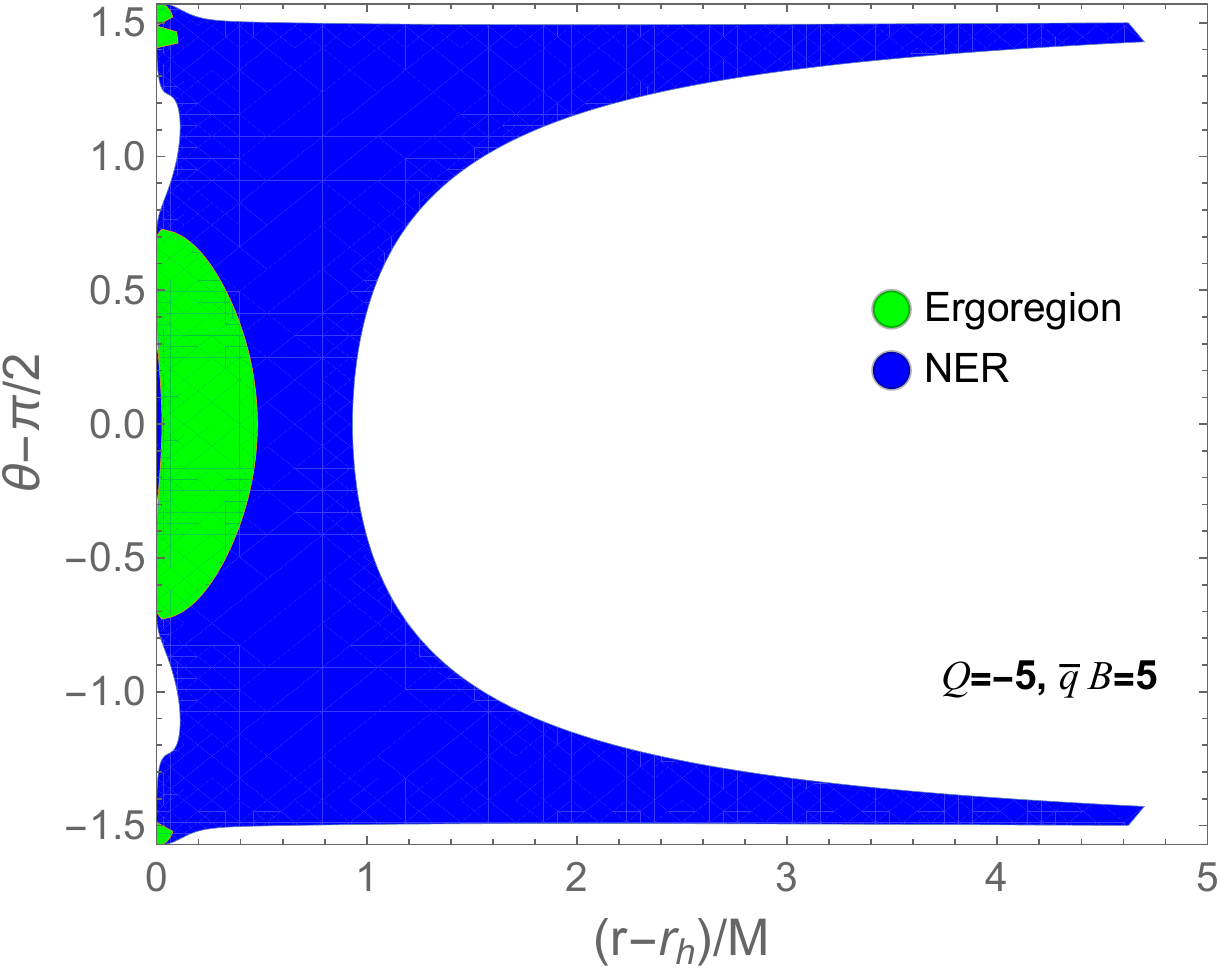}\quad
        \includegraphics[width=0.31\linewidth]{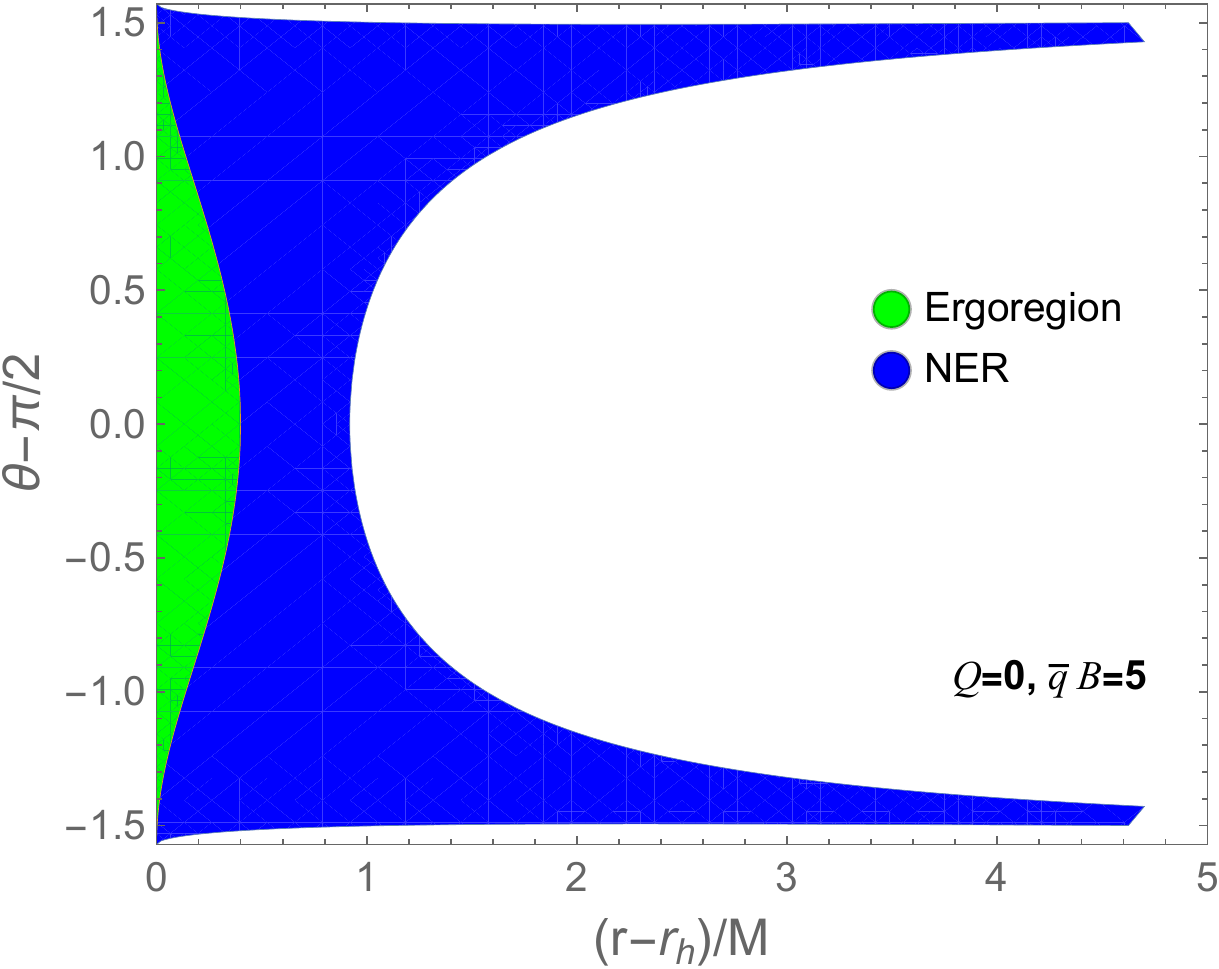}\quad
        \includegraphics[width=0.31\linewidth]{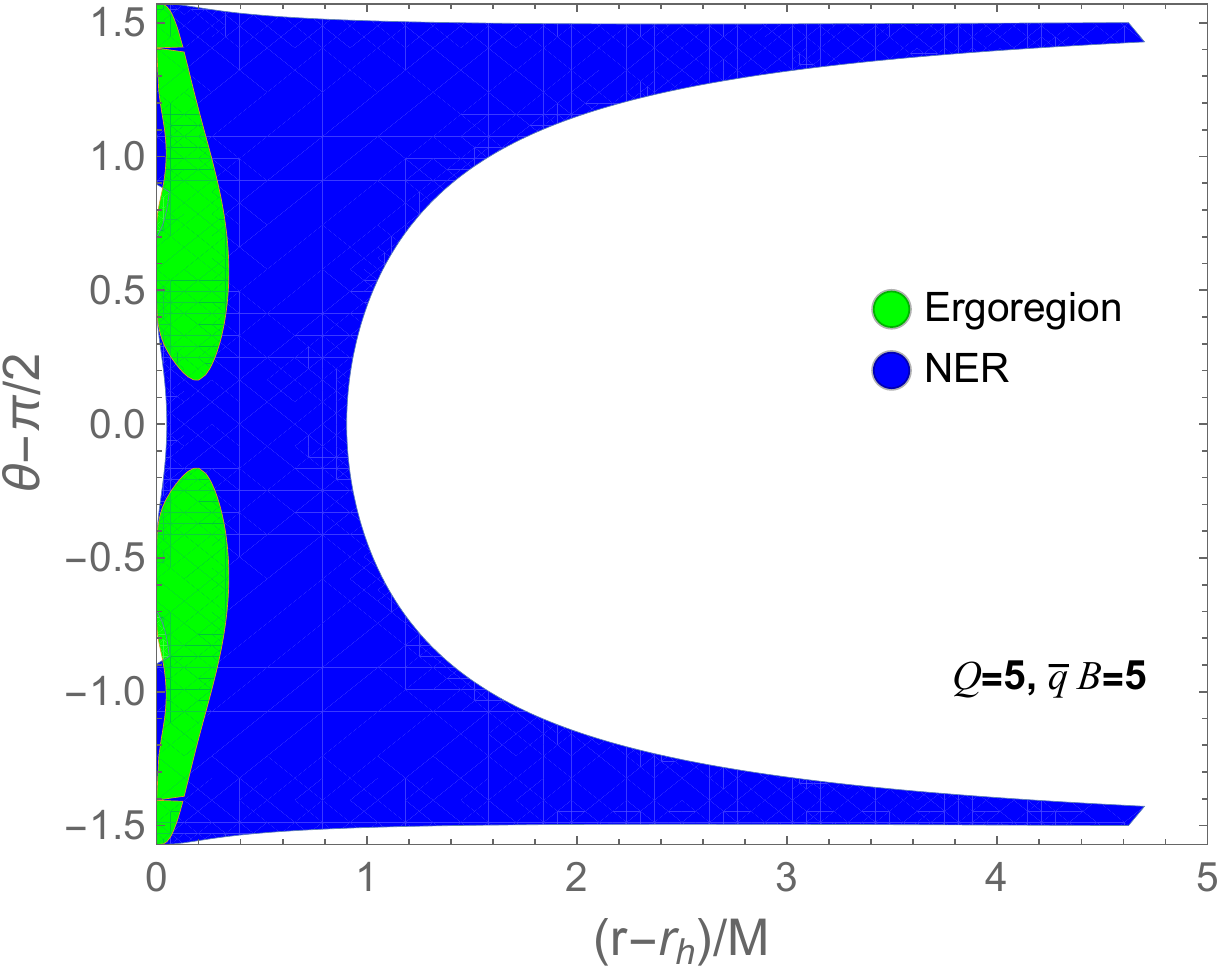}
        \caption{Shapes of the negative-energy region (NER) and ergoregion for different ${\cal Q}$ with $\bar{q} B >0$. We set $a = 0.8$, and $\ell=-1$ (upper and middle rows), $1$ (bottom row). In plotting the NER, the CTCs region has been excluded. }
	\label{NERQ-2}
\end{figure}

\begin{figure}[htb]
	\centering
        \includegraphics[width=0.31\linewidth]{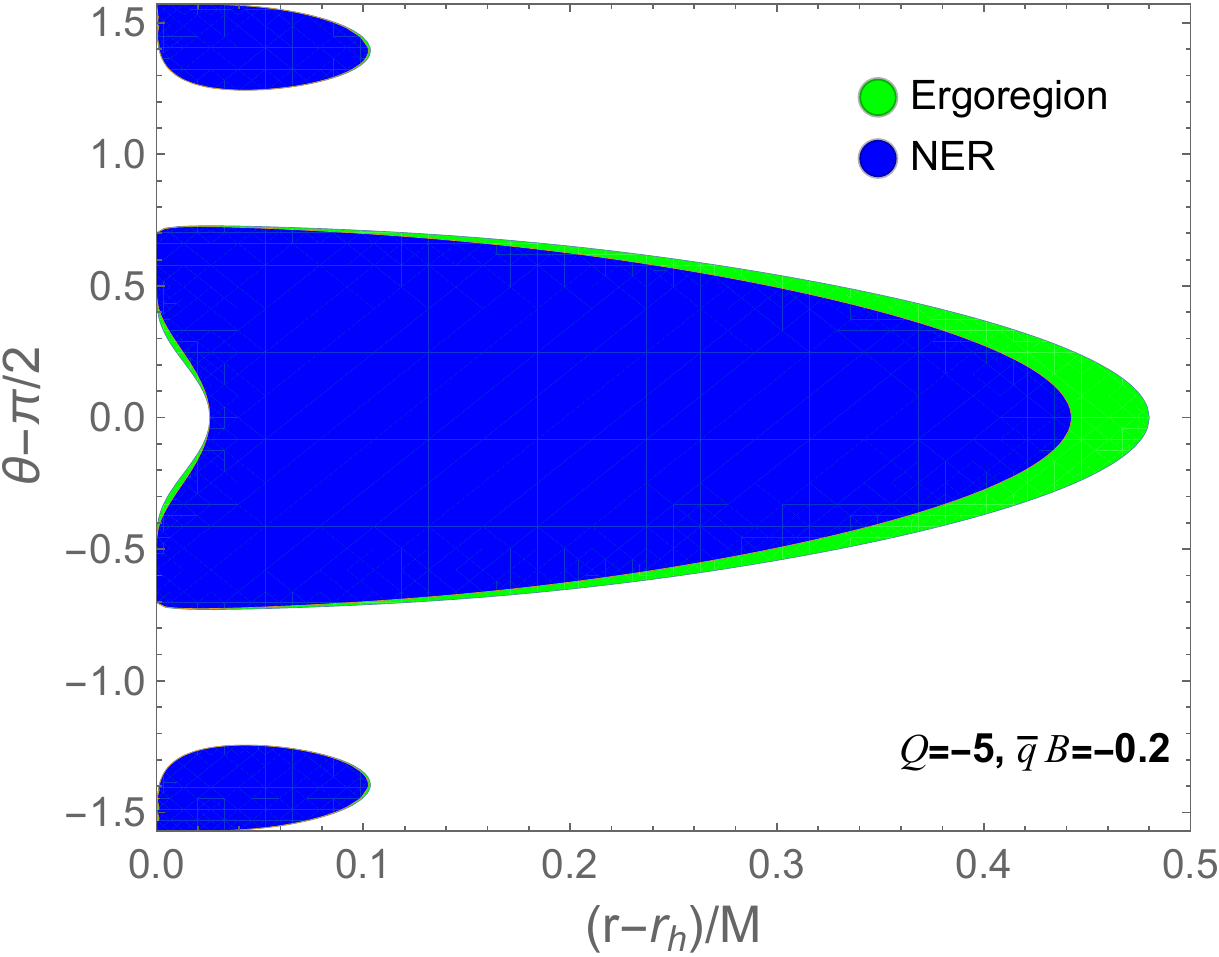}\quad
        \includegraphics[width=0.31\linewidth]{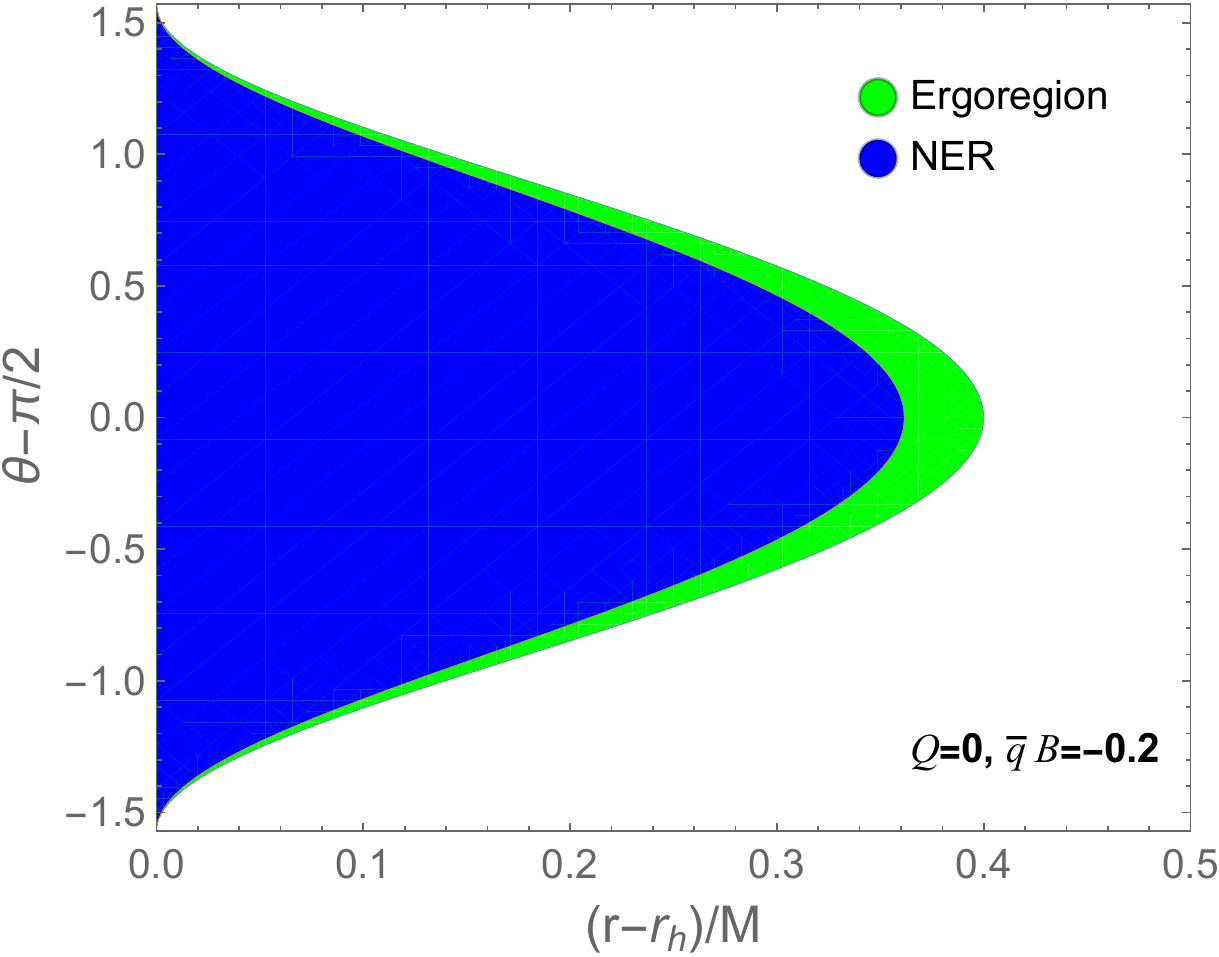}\quad
        \includegraphics[width=0.31\linewidth]{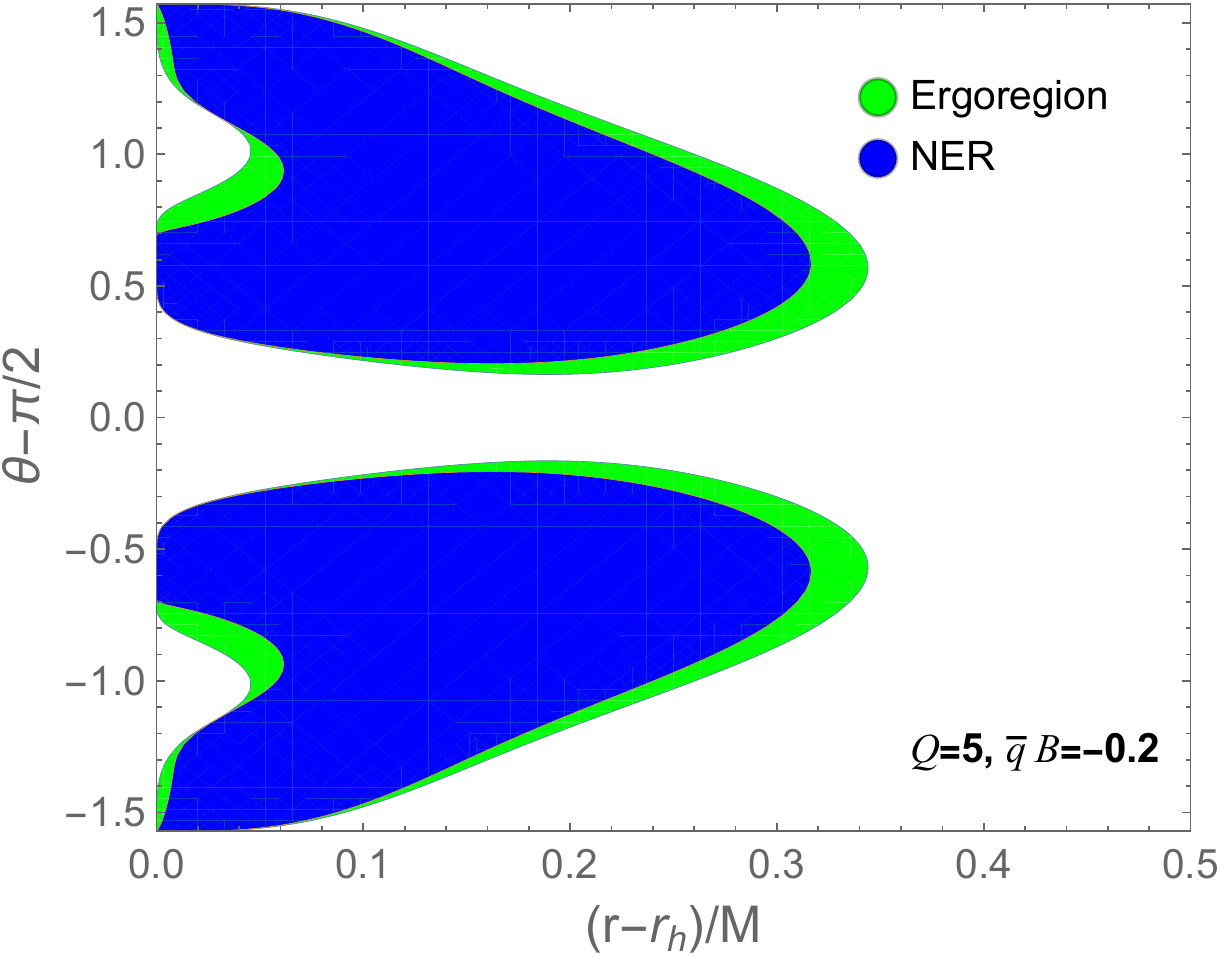}\\
       \vspace{0.5cm}
       
        \includegraphics[width=0.31\linewidth]{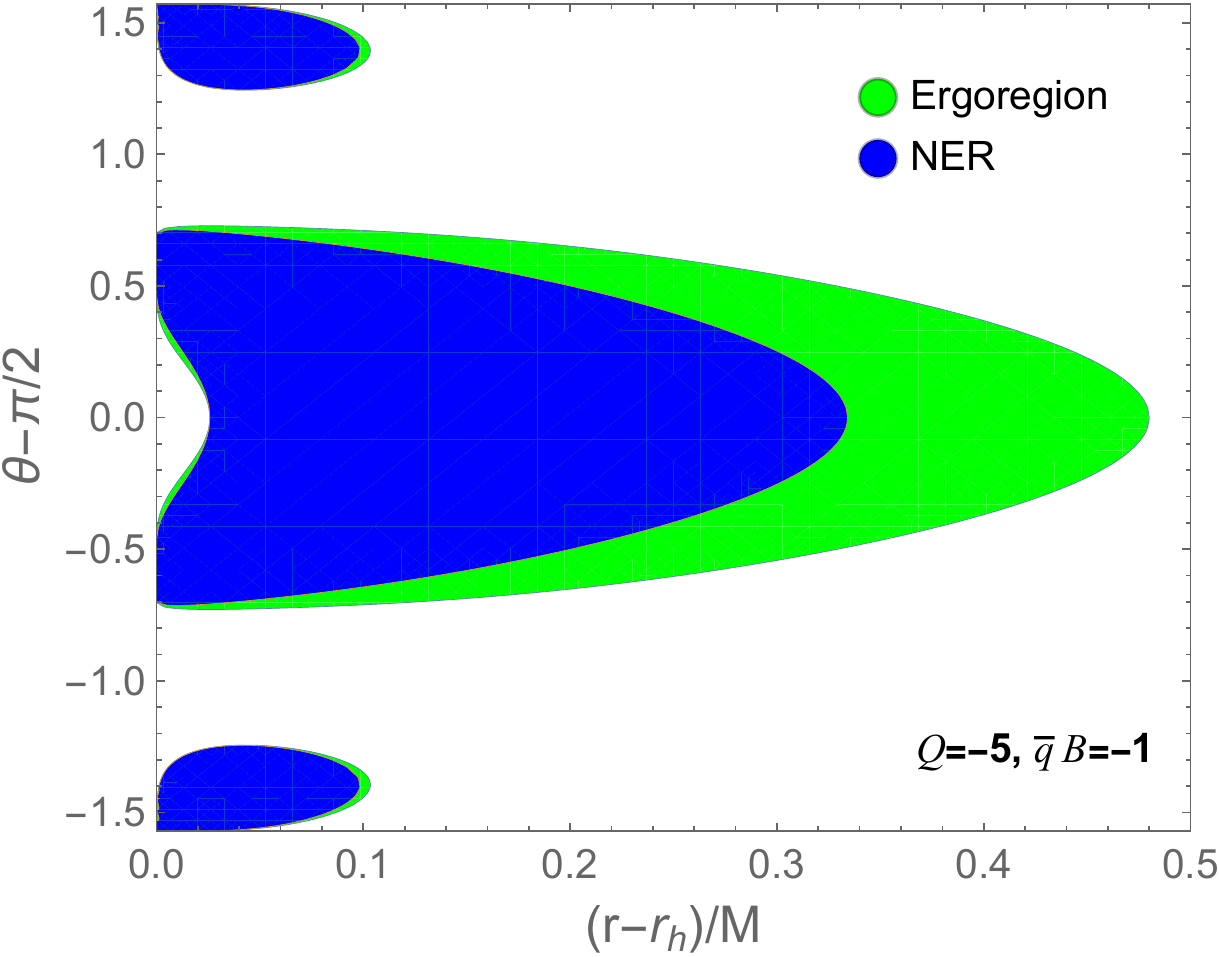}\quad
        \includegraphics[width=0.31\linewidth]{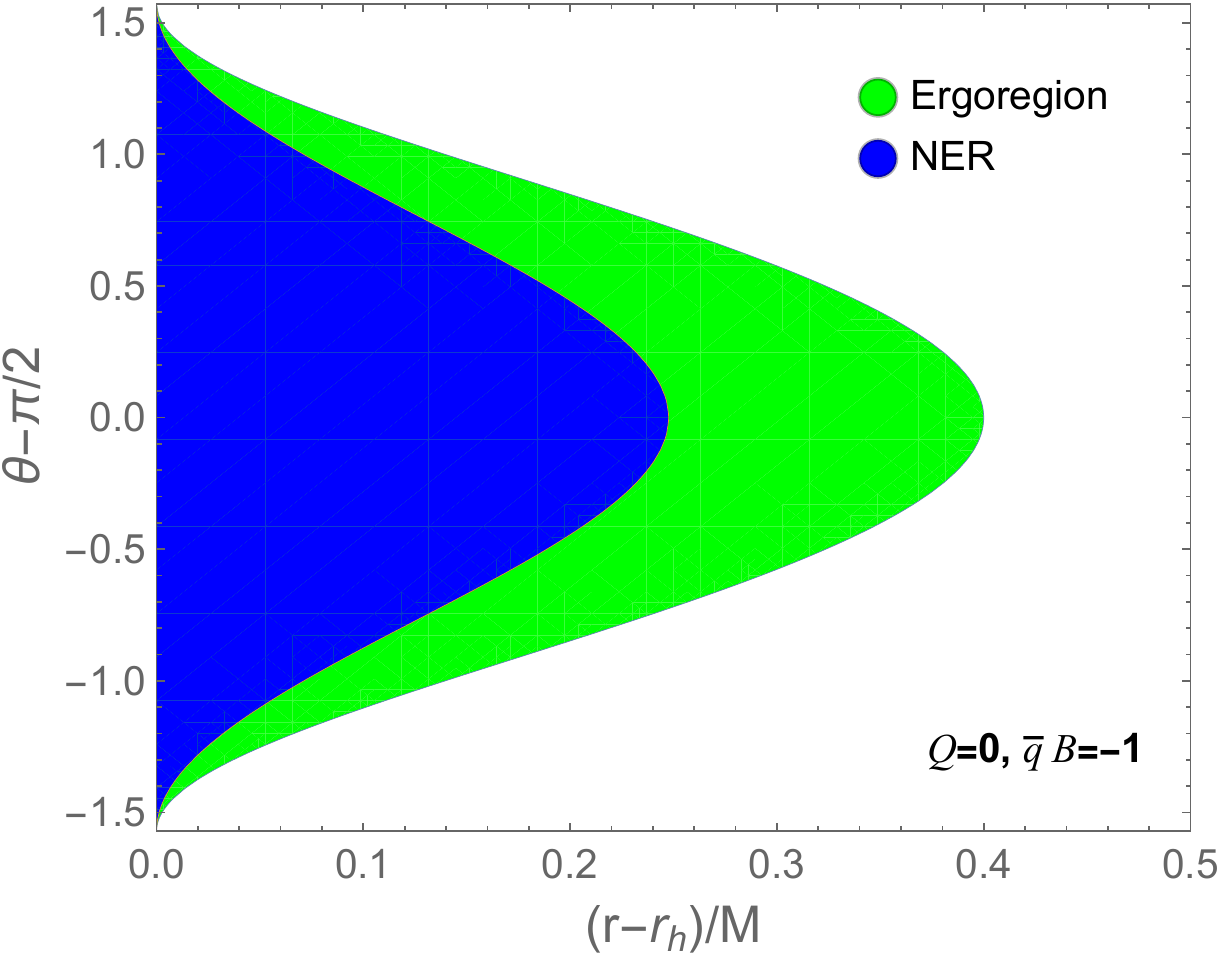}\quad
        \includegraphics[width=0.31\linewidth]{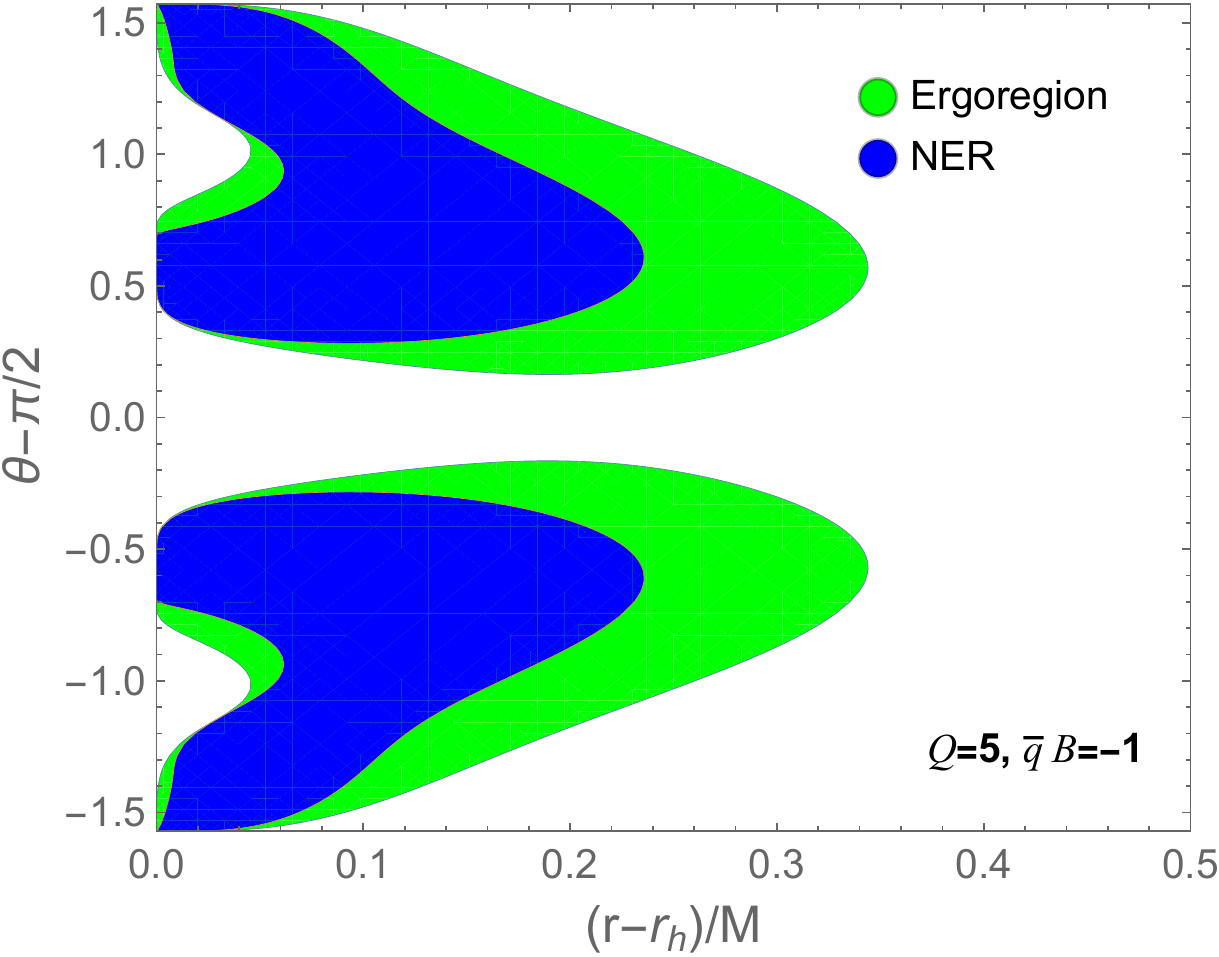}\\
        \vspace{0.5cm}
        \caption{Shapes of the negative-energy region (NER) and ergoregion for different ${\cal Q}$ with $\bar{q} B <0$. We set $a = 0.8$ and $\ell=-10$. In plotting the NER, the CTCs region has been excluded.}
	\label{NERQ-3}
\end{figure}

In Figs. \ref{NERQ-1}, \ref{NERQ-2} and \ref{NERQ-3}, the shape of NER is plotted for three cases: $\bar{q} B =0, \bar{q} B >0$, and $\bar{q} B <0$. The shape of the ergoregion is also shown for comparison.

Fig. \ref{NERQ-1} illustrates the case where $\bar{q} B =0$. In this scenario, $V_{\rm eff} (r)<0$ is possible only if $\ell<0$, and the NER is always contained within the ergoregion. As $\ell$ becomes more negative, the NER expands, and the zero-energy surface approaches the ergosurface. From the figure, it can also be seen that ${\cal Q}$ significantly influences the near-horizon geometry, leading to a multi-lobe structure of the NER when $|{\cal Q}|$ is sufficiently large, similar to its effect on the ergoregion and CTCs region (Fig. \ref{ErgoregionCTCs}). 

Fig. \ref{NERQ-2} illustrates the case where $\bar{q} B >0$. From the figure, it can be seen that even for small values of $\bar{q} B$ (e.g., $\bar{q} B = 0.2$ as shown in the upper row), the NER may extend beyond the ergoregion in certain angular directions. As $\bar{q} B$ increases, the NER continues to expand and eventually fully contains the ergoregion once $\bar{q} B$ exceeds a specific value. For ${\cal Q} \neq 0$ and small $\bar{q} B$, both regions exhibit a multi-lobe structure, similar to the $\bar{q} B = 0$ case. However, as $\bar{q} B$ grows, the NER gradually reverts to a simply connected structure, as shown in the middle and bottom rows. When $\ell$ becomes more negative, the NER may either expand or shrink, depending on the value of $\bar{q} B$. Furthermore, in this case, $V_{\rm eff} (r)<0$ is also possible for $\ell>0$, as shown in the bottom row of the figure. For $\ell>0$, the NER extends far beyond the ergoregion, particularly along angular directions near the poles ($\theta = 0, \pi$).

Fig. \ref{NERQ-3} illustrates the case where $\bar{q} B <0$. In this scenario, the NER exists only if $\ell<0$ and is always contained within the ergoregion. The more negative $\ell$ becomes, the larger the NER. Comparing the two rows, it can be seen that as $|\bar{q} B|$ increases, the NER shrinks and eventually disappears when $\bar{q} B$ is sufficiently negative.

\section{Energy extraction through magnetic Penrose process}

\subsection{General formalism}

We now consider energy extraction from the body via the magnetic Penrose process (MPP) \cite{dhurandharEnergyextractionProcessesKerr1984,dhurandharEnergyextractionProcessesKerr1984a,1985ApJ...290...12W,parthasarathyHighEfficiencyPenrose1986,Bhat:1985hpc,Wagh:1989zqa,tursunovFiftyYearsEnergy2019,Stuchlik:2021unj}. For simplicity, we examine a straightforward scenario: a particle $0$ moving in the equatorial plane splits into two fragments $1$ and $2$, which also move in the equatorial plane. In this process, 4-momentum and electric charge are conserved, so we have
\begin{align}
 m_0 U^\mu_0 &= m_1 U^\mu_1 + m_2 U^\mu_2, \label{MomentumConservation}\\
 q_0 & = q_1 + q_2. \label{ChargeConservation}
\end{align}
The subscripts $0, 1, 2$ label the three particles. Combining the above two conservation equations yields the conservation laws for energy and angular momentum
\begin{align}
    E_0 &= E_1 + E_2,\label{EnergyConservation}\\
    L_0 &= L_1 + L_2.\label{AngularMomentumConservation}
\end{align}
Energy extraction occurs when particle $1$ acquires negative energy ($E_1<0$) and falls into the body, while particle $2$ escapes to infinity with excess energy ($E_2 > E_0$) to conserve the total energy. 

Denoting $U^\mu_i = \dot{t}_i (1, v_i, 0, \Omega_i) (i=0,1,2)$, the three components of the 4-momentum conservation equation (\ref{MomentumConservation}) can be written as
\begin{align}
    m_0 \dot{t}_0 &= m_1 \dot{t}_1 + m_2 \dot{t}_2,\label{TimeMomentumConservation}\\
    m_0 v_0 \dot{t}_0 &= m_1 v_1 \dot{t}_1 + m_2 v_2 \dot{t}_2,\label{RadialMomentumConservation}\\
    m_0 \Omega_0 \dot{t}_0 &= m_1 \Omega_1 \dot{t}_1 + m_2 \Omega_2\dot{t}_2.\label{PhiMomentumConservation}
\end{align}
Also, we have the relation from (\ref{Energy})
\begin{align}
    E_i = - m_i \dot{t}_i X_i - q_i A_t,\qquad X_i \equiv g_{tt} + \Omega_i g_{t\phi}. \label{EnergyTimeMomentumRelation}
\end{align}

The efficiency of the energy extraction is defined as
\begin{align}
    \eta \equiv \frac{E_2 - E_0}{E_0}. \label{Efficiency-1}
\end{align}
With Eqs. (\ref{TimeMomentumConservation}), (\ref{PhiMomentumConservation}) (or (\ref{RadialMomentumConservation})) and (\ref{EnergyTimeMomentumRelation}), the energy of particle 2 can be expressed as
\begin{align}
    E_2 = \kappa (E_0 + q_0 A_t) - q_2 A_t,
\end{align}
where
\begin{align}
    \kappa &= \frac{\Omega_0 - \Omega_1}{\Omega_2 - \Omega_1} \frac{X_2}{X_0} \\
         &= \frac{v_0 X_1 - v_1 X_0}{v_2 X_1 - v_1 X_2} \frac{X_2}{X_0}.
\end{align}
With this equation, the efficiency (\ref{Efficiency-1}) can be written as
\begin{align}
    \eta = \kappa -1 + \kappa \frac{\bar{q}_0 A_t}{e_0} - \frac{m_2}{m_0}\frac{\bar{q}_2 A_t}{e_0} . \label{Efficiency-2}
\end{align}
This is the most general formalism for the energy extraction efficiency via the MPP. To more clearly illustrate how parameters affect the efficiency, we introduce the following reasonable approximations. First, following \cite{parthasarathyHighEfficiencyPenrose1986}, we set $q_0 =0$, such that  $q_1 = - q_2 \equiv q$. This assumption corresponds to the extremely efficient regime of MPP \cite{Stuchlik:2021unj} and is well-justified, as astronomical objects like stars generally carry no net charge. Additionally, we take $e_0 \approx 1$, considering that in nearly all realistic scenarios, the incident particle initially moves non-relativistically \cite{parthasarathyHighEfficiencyPenrose1986}. Under this approximation, the angular velocity of particle $0$ can be derived using (\ref{AngularVelocity}) (\ref{EnergyTimeMomentumRelation})
\begin{align}
    \Omega_0 = \frac{-g_{t\phi} (1 + g_{tt}) + \sqrt{(g_{t\phi}^2 - g_{tt} g_{\phi\phi}) (1 + g_{tt})}}{g_{\phi\phi} + g_{t\phi}^2}.
\end{align}
The positive sign for the radical has been taken considering the frame-dragging effect of the spacetime. 

Furthermore, we aim to explore the possible maximum efficiency. As argued in \cite{parthasarathyHighEfficiencyPenrose1986}, $\kappa$ (and thus $\eta$) is maximized when all radial velocities vanish(i.e., $v_i = 0 (i=0,1,2)$) and the angular velocities of the two fragments take their limiting values (i.e., $\Omega_1 = \Omega_-$ and $\Omega_2 = \Omega_+$). In this case, the efficiency becomes
\begin{align}
    \eta = \frac{1}{2} \left[\sqrt{1+g_{tt}} - 1 \right] +  \hat{\bar{q}} A_t, \label{Efficiency-3}
\end{align}
where $\hat{\bar{q}} \equiv \frac{m_1}{m_0} \bar{q}$ may be referred to as the rescaled charge-to-mass ratio of particle $1$. From (\ref{Efficiency-3}), it is evident that under the above approximations, $\eta$ depends only on four parameters: $\{a, {\cal Q}, \hat{\bar{q}} B\}$, as well as the splitting point $r=r_\ast$. Additionally, the mass ratio $m_1/m_0$ is constrained by
\begin{align}
    m_1 + m_2 \leq m_0, \label{MassConstrain}
\end{align}
which can be understood from the 4-momentum conservation in a local rest frame of particle $0$. 

Some initial insights can be gleaned from (\ref{Efficiency-3}). There are two terms contributing to $\eta$: the first term is a pure geometric factor, while the second arises from the electromagnetic interaction. In the extreme Kerr limit (${\cal Q} =  0, a = M$) without the magnetic field ($B = 0$), only the first term contributes, and the efficiency $\eta$ reaches the maximum value as $r_\ast \rightarrow r_h$, i.e., $\eta_{\rm max}\approx 20.7\%$, in accordance with the known result for the mechanical Penrose process \cite{Penrose:1971uk, 1983mtbh.book.....C}. When the electromagnetic interaction is incorporated, the second term dominates if $|\hat{\bar{q}} B| \gtrsim 1$, and can cause $\eta$ to exceed $100\%$ for sufficiently large $|\hat{\bar{q}} B|$, as demonstrated below.  

Given the intricacy of the metric, the complete expression for efficiency (\ref{Efficiency-3}) is rather involved. For a slowly rotating ($a \ll 1$) and slightly deformed ($|{\cal Q}| \ll 1$) body, it simplifies to
\begin{align}
    \eta = \frac{1}{2} \left(\sqrt{\frac{2}{r_\ast}} - 1\right) - \frac{r_\ast-2}{16 \sqrt{2 r_\ast}} \left[6 - 6 r_\ast + (3 r_\ast^2 - 6 r_\ast +2) \ln\left(\frac{r_\ast}{r_\ast-2}\right)\right] {\cal Q} + \frac{\hat{\bar{q}} B}{r_\ast} a + {\cal O}(a^2, {\cal Q}^2, a {\cal Q}). \label{Efficiency-4}
\end{align}
Considering that the event horizon satisfies $r_h = 2 + {\cal O} (a^2)$ and the splitting point satisfies $r_\ast > r_h$ in this context, we make the following observations: (i) A more positive $\hat{\bar{q}} B$ value yields a higher $\eta$, suggesting a more advantageous condition for energy extraction; (ii) For a small but larger $a$, $\eta$ increases when $\hat{\bar{q}} B >0$, but decreases when $\hat{\bar{q}} B <0$; (iii) Because the coefficient of the ${\cal Q}$-term is consistently negative, a small but more negative ${\cal Q}$ results in a higher $\eta$.

\subsection{Efficiency of energy extraction}

In this subsection, using the exact expression for efficiency given by Eq. (\ref{Efficiency-3}), we analyze the influence of various parameters on efficiency via numerical calculations. As discussed above, the would-be event horizon $r=r_h$ contains a naked singularity when ${\cal Q} \neq 0$, and the CTCs region (with outer radius $r=r_c$) emerges adjacent to the horizon if $|{\cal Q}|$ exceeds a critical value. The general formalism of the NER and efficiency becomes inapplicable as the splitting point $r_\ast$ approaches these regions. To exclude such unphysical regions, we follow the approach used in \cite{Bambi:2019tjh,Chakraborty:2024aug} for handling superspinars, setting the radius of the compact object as $r_s = r_c(1+10^{-3})$ to obscure these unphysical regions. We thus consider the splitting point $r_\ast$ to lie outside the surface of the compact object, i.e., $r \geq r_s$.

\begin{figure}[htb]
	\centering
        \includegraphics[width=0.45\linewidth]{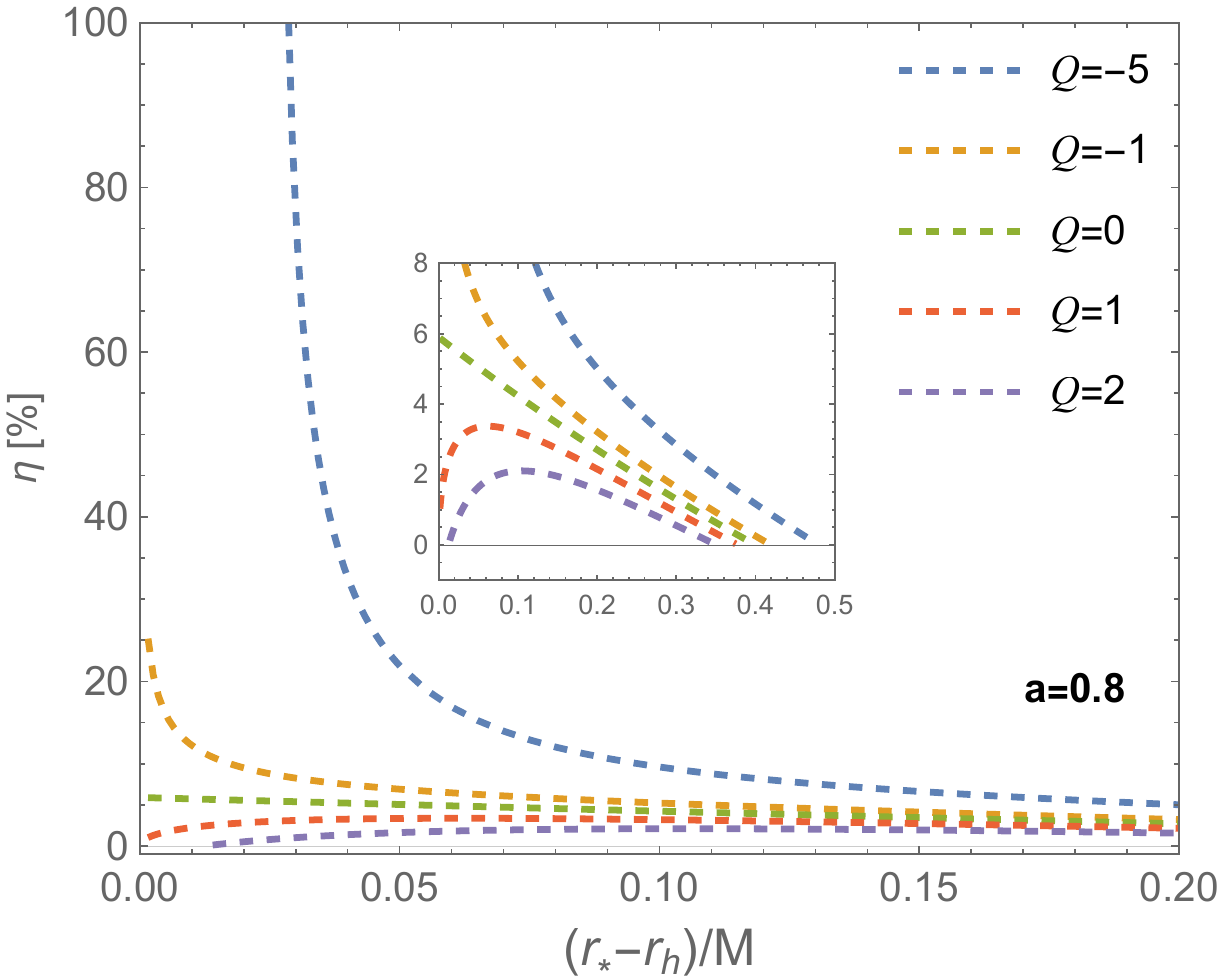}\quad
        \includegraphics[width=0.45\linewidth]{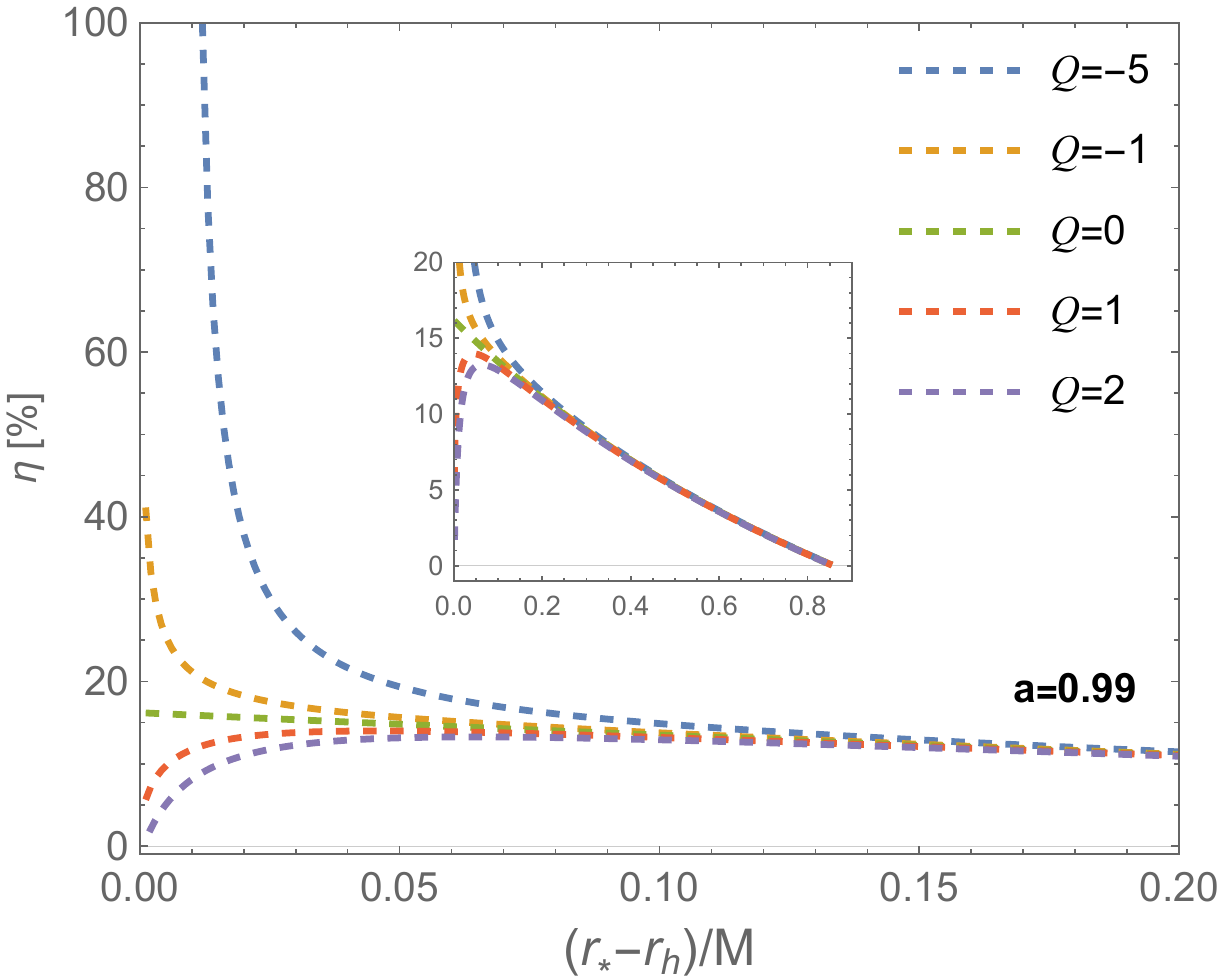}
        
        \caption{Efficiency of energy extraction $\eta$ as a function of the splitting point $r_\ast$ for various ${\cal Q}$ with $B=0$. }
	\label{Fig: Efficiency-1}
\end{figure}

Even in the absence of the magnetic field ($B=0$), the efficiency $\eta$ deviates from its Kerr value due to the anomalous quadrupole moment ${\cal Q}$. In this scenario, only the purely geometric term contributes, such that $\eta = \frac{1}{2} \left[\sqrt{1+g_{tt}} - 1 \right]$, which depends on three parameters: $\{a, {\cal Q}, r_\ast \}$. Notably, $\eta>0$ only if the splitting point lies within the ergoregion, i.e., $r_\ast \leq r_e$. As shown in Fig.  \ref{Fig: Efficiency-1}, for a fixed $r_\ast$, $\eta$ increases as ${\cal Q}$ becomes more negative. This indicates that a more negative ${\cal Q}$ enhances energy extraction efficiency. For ${\cal Q} \leq 0$, $\eta$ exhibits a monotonically decreasing behavior with respect to $r_\ast$. As $r_\ast$ approaches the object's surface $r_s$, $\eta$ increases rapidly and reaches a maximum value $\eta_{\rm max}$ at $r=r_s$. If ${\cal Q}$ is sufficently negative,   $\eta_{\rm max}$ can even exceed $100\%$. In contrast, when ${\cal Q}>0$, $\eta$ first increases and then decreases as $r_\ast$ decreases.

\begin{figure}[htb]
	\centering
        \includegraphics[width=0.6\linewidth]{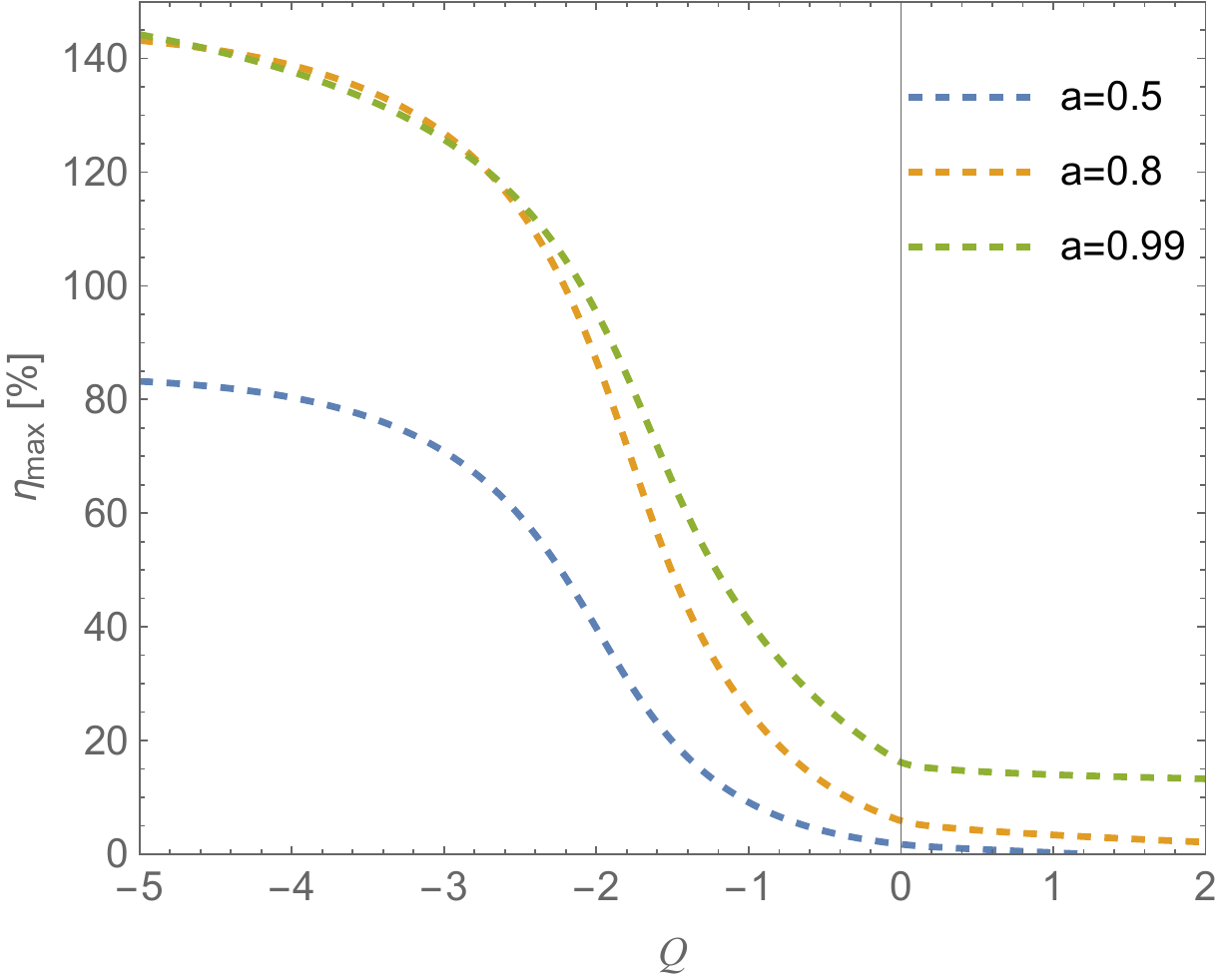}
                
        \caption{Maximal efficiency of energy extraction $\eta_{\rm max}$ as a function of ${\cal Q}$ for $B=0$.}
	\label{Fig: Efficiency-2}
\end{figure}

In Fig. \ref{Fig: Efficiency-2}, the maximal efficiency $\eta_{\rm max}$ under $B=0$ is depicted as a function of ${\cal Q}$ for different values of the spin $a$. The figure reinforces the earlier observation that for a fixed $a$, a more negative ${\cal Q}$ results in a larger $\eta_{\rm max}$.  Within the range of ${\cal Q}$ considered in our study, the possible $\eta_{\rm max}$ is on the order of $100\%$. As ${\cal Q}$ transitions from positive to negative, $\eta_{\rm max}$ undergoes three distinct evolutionary stages: an initially slow growth, followed by a rapid increase, and finally a return to slow growth. For a fixed ${\cal Q}$, a higher spin $a$ generally leads to a larger $\eta_{\rm max}$. However, in the high-spin regime, when ${\cal Q}$ is sufficiently negative, the influence of ${\cal Q}$  dominates over that of the spin $a$, rendering $\eta_{\rm max}$ nearly independent of $a$.

\begin{figure}[htb]
	\centering
        \includegraphics[width=0.45\linewidth]{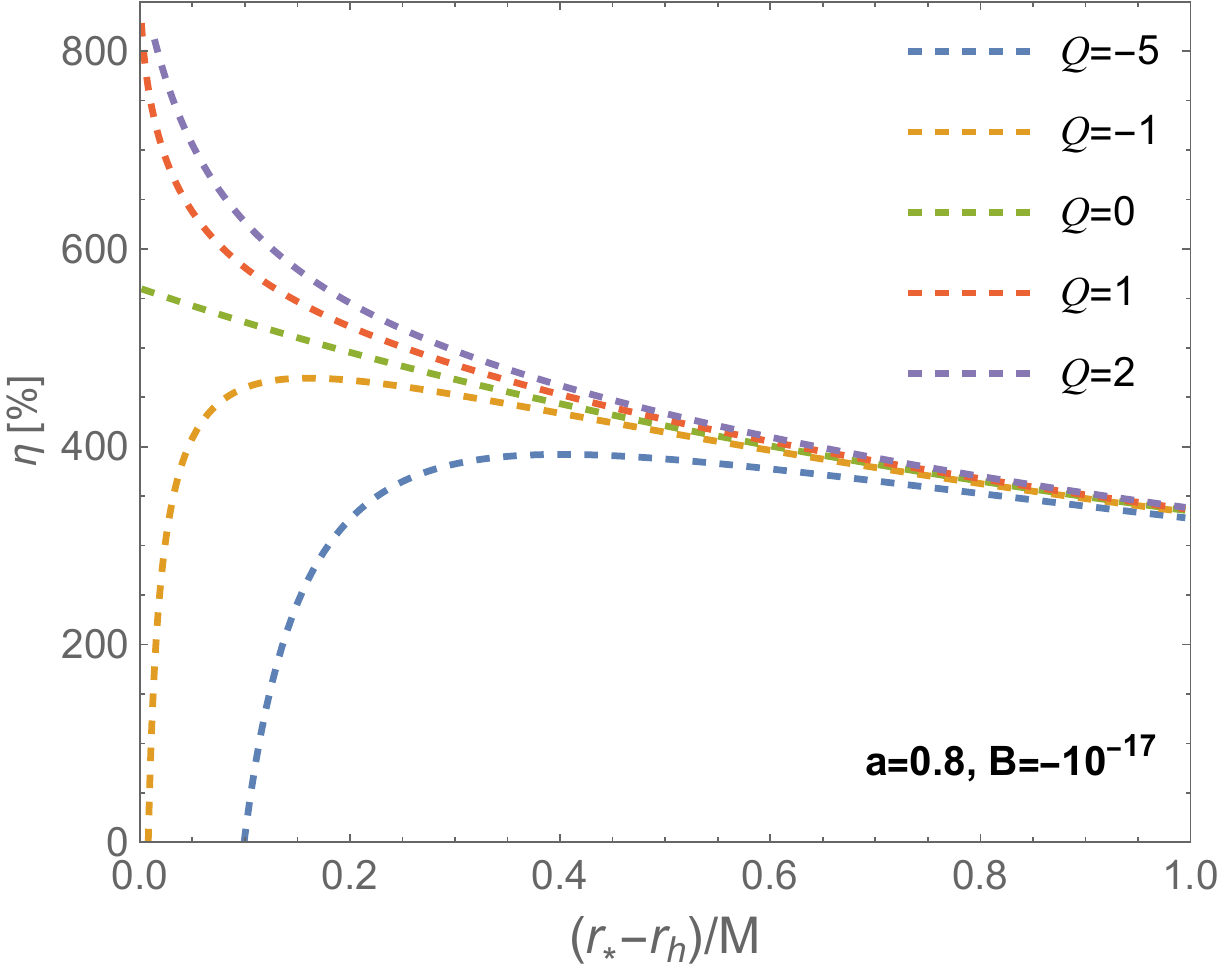}\quad
        \includegraphics[width=0.45\linewidth]{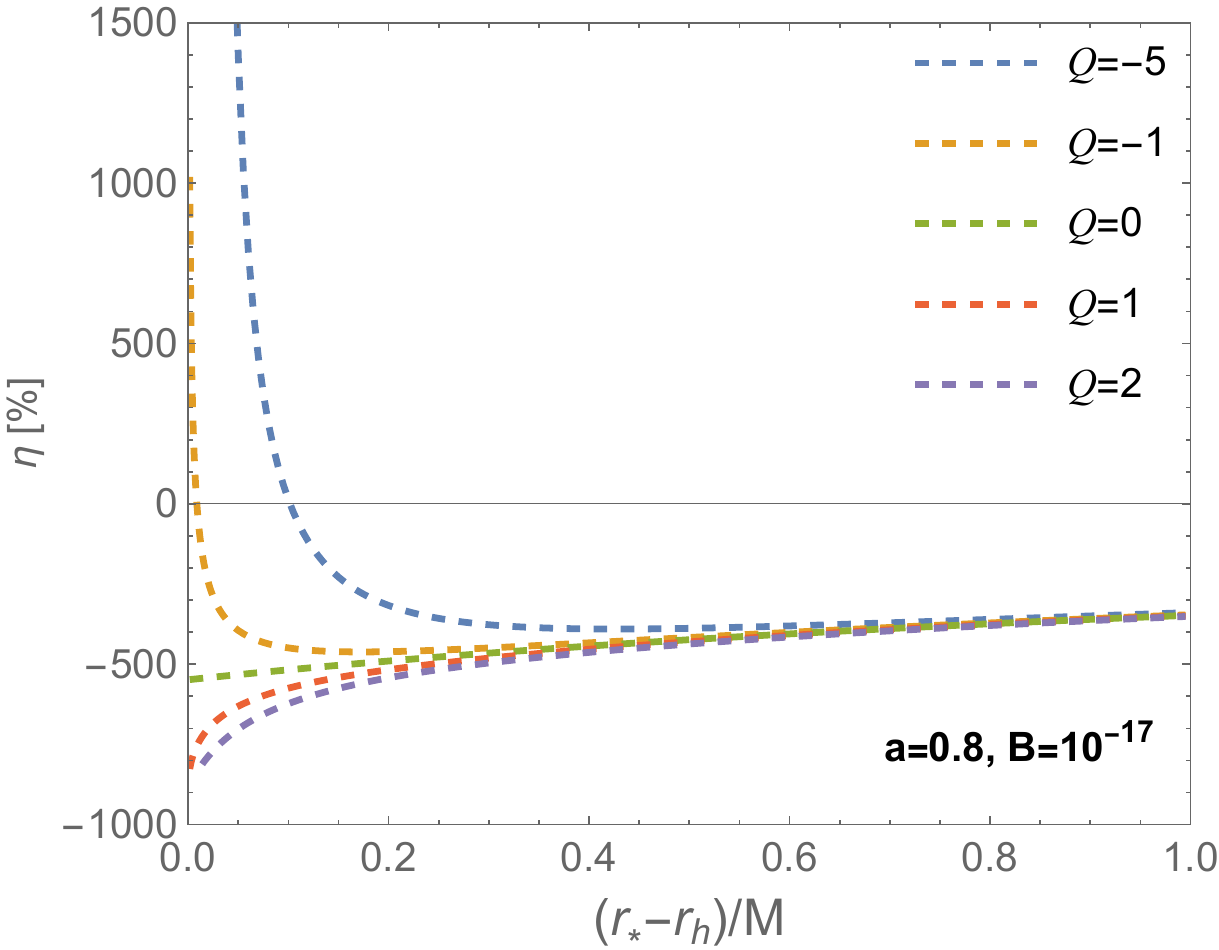}
        
        \caption{Efficiency of energy extraction $\eta$ as a function of the splitting point $r_\ast$ for various ${\cal Q}$ and $B \neq 0$. }
	\label{Fig: Efficiency-3}
\end{figure}

When the electromagnetic interaction is incorporated, the scenario changes significantly. Without loss of generality, we follow \cite{tursunovFiftyYearsEnergy2019} to analyze a specific process: the beta decay of a neutron (particle $0$) into an electron $e^-$ (particle $1$), a proton $p^+$ (particle $2$), and an anti-neutrino $\bar{\nu}_e$, described by
\begin{align}
    n \rightarrow e^{-} + p^+ + \bar{\nu}_e. \label{BetaDecay}
\end{align}
The influence of the anti-neutrino on the proton's energy is neglected here. In our units, $m_1/m_0 \sim 1/1840$ and $\bar{q} = q_e/m_e \sim - 2.04 \times 10^{21}$. With the rescaled charge-to-mass ratio reaching an extremely high value $\hat{\bar{q}} \sim -1.1 \times 10^{18}$, the efficiency defined in (\ref{Efficiency-3}) is entirely dominated by the electromagnetic term for $|B| > 10^{-18}$, as illustrated in Fig. \ref{Fig: Efficiency-3} where $|B| = 10^{-17}$. In physical units, this corresponds to $|B| \sim 2.35 \times 10^{2} {\rm Gauss}$ for $M=M_\odot$, $|B| \sim 2.35 \times 10^{-4} {\rm Gauss}$ for $M=10^6 M_\odot$, or $|B| \sim 2.35 \times 10^{-8} {\rm Gauss}$ for $M=10^{10} M_\odot$. 

From the left panel of Fig. \ref{Fig: Efficiency-3} (with $B<0$), it is evident that for a fixed $r_\ast$, $\eta$ increases as ${\cal Q}$ becomes more positive, which is in stark contrast to the $B=0$ case depicted in Fig. \ref{Fig: Efficiency-1}. For a fixed ${\cal Q}>0$, $\eta$ rises as $r_\ast$ decreases, reaching a maximum value $\eta_{\rm max}$ as $r_\ast \rightarrow r_s$, where $\eta_{\rm max}$ is typically on the order of $500\%$. Conversely, for a fixed ${\cal Q}<0$,  $\eta$ first increases and then decreases as $r_\ast$ decreases. A notable observation, when compared to the $B=0$ case (Fig. \ref{Fig: Efficiency-1}), is that $\eta$ maintains a high value even when the splitting point $r_\ast$ lies far beyond the ergoregion (yet still within the NER).

The case with $B>0$ (right panel of Fig. \ref{Fig: Efficiency-3}) exhibits stark differences from the $B<0$ case. Here, a more negative ${\cal Q}$ leads to a larger $\eta$. Notably, $\eta<0$ when ${\cal Q} \geq 0$, and $\eta$ becomes positive only if ${\cal Q} <0$ and $r_\ast$ is in close proximity to $r_s$.

\begin{figure}[htb]
	\centering
        \includegraphics[width=0.45\linewidth]{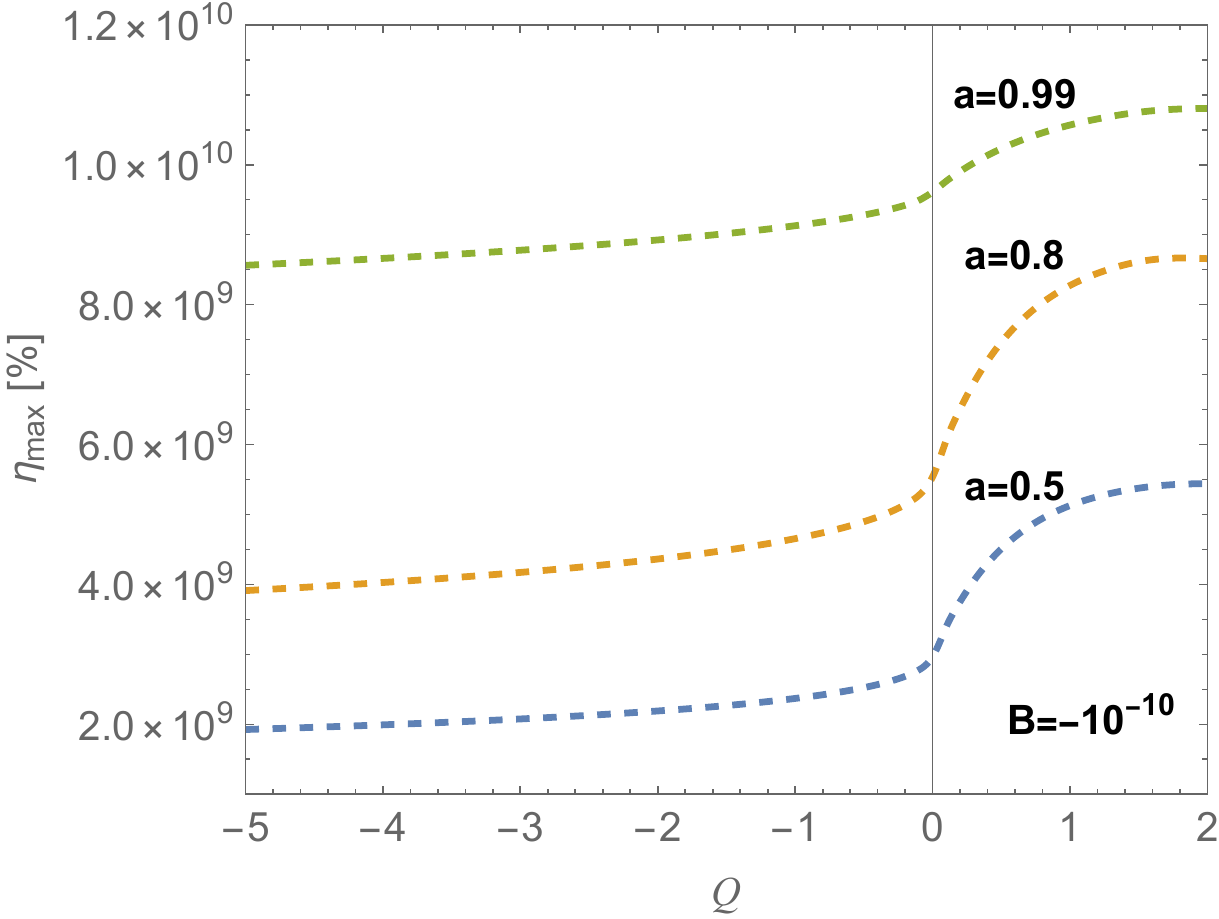}\quad
        \includegraphics[width=0.45\linewidth]{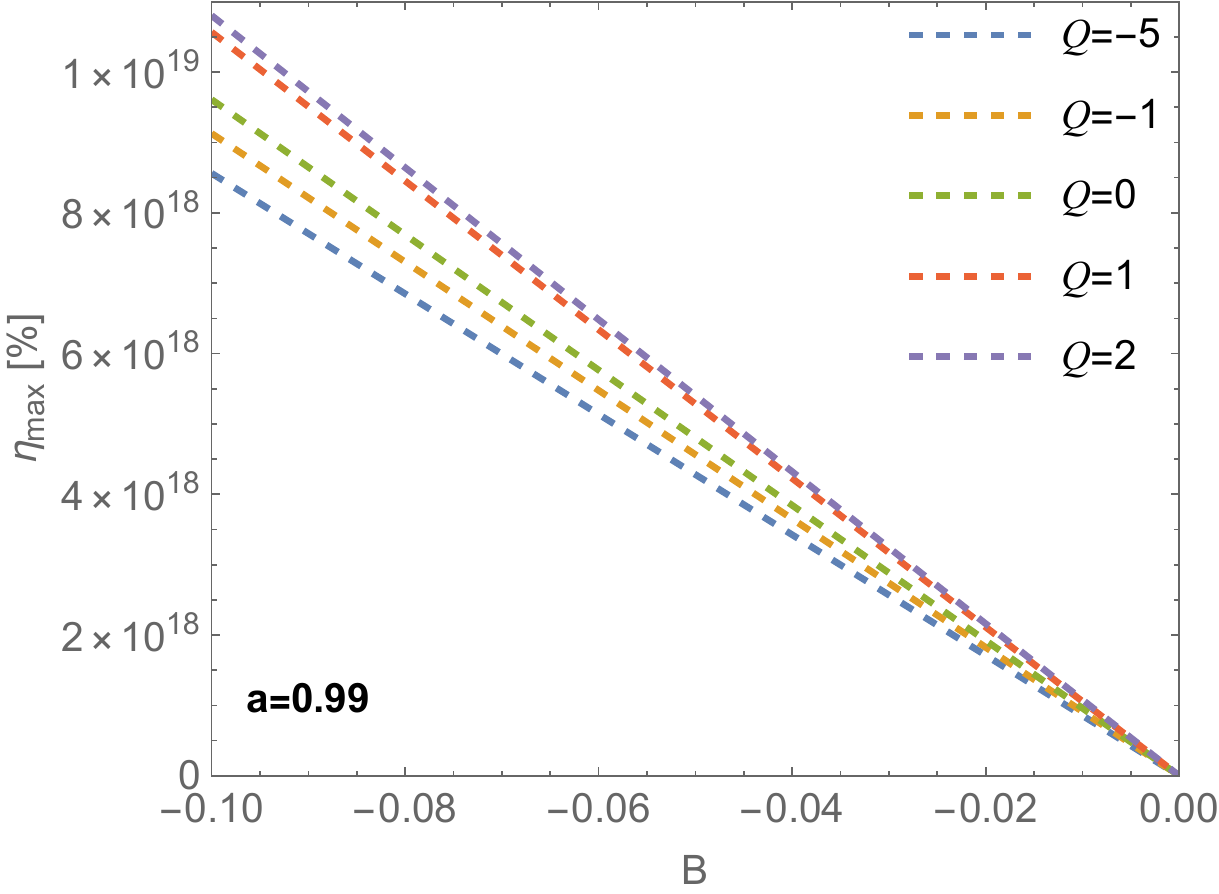}
                
        \caption{Maximal efficiency of energy extraction $\eta_{\rm max}$ as a function of ${\cal Q}$ (left panel) and $B$ (right panel).}
	\label{Fig: Efficiency-4}
\end{figure}

The linear dependence of $\eta$ on $B$ implies that the order of the magnitude of $\eta$ scales with $|B|$, as demonstrated in Fig. \ref{Fig: Efficiency-4}. In the left panel, with $B=-10^{-10}$, the product $\hat{\bar{q}} B \sim 1.1 \times 10^8$. Depending on the values of $\{a, {\cal Q} \}$, $\eta_{\rm max}$ reaches an order of magnitude of $10^{9-10} \%$. When $B$ increases to the order of $0.1$,  $\eta_{\rm max}$ can even reach $10^{18-19} \%$. A larger $\eta_{\rm max}$ is observed for more positive ${\cal Q}$ or higher spin $a$, a trend also evident in the right panel. Similar to the $B=0$ case, $\eta_{\rm max}$ undergoes three distinct evolutionary stages as ${\cal Q}$ varies from negative to positive values. For fixed $a$ and $B$, $\eta_{\rm max}$ increases to roughly $1.5$ to $2$ times its original value as ${\cal Q}$ varies from $-5$ to $2$. For high spins (e.g., $a=0.99$), $\eta_{\rm max}$ drops by many orders of magnitude or even turns negative when $B$ is positive with the same magnitude, rendering the $B>0$ portion invisible in the right panel. This suggests that energy extraction is more favorable when $q$ and $B$ share the same sign. 

\section{Summary and discussions}

In this work, we investigate the MPP in the QM spacetime \cite{1985PhLA..109...13Q}, which is a non-Kerr rotating spacetime with an anomalous quadrupole moment. Despite exhibiting pathologies like naked singularities and closed time-like curves (CTCs), it can be employed to describe the exterior gravitational field of a rotating body with an arbitrary quadrupole moment. Additionally, we consider the body immersed in an asymptotically uniform magnetic field, whose configuration is characterized by Wald's solution \cite{Wald:1974np}. 

The anomalous quadrupole moment, parameterized by ${\cal Q}$, significantly modifies the near-horizon geometry, as illustrated in Fig. \ref{ErgoregionCTCs}.  It is observed that when $|Q|$ exceeds a critical value, a CTCs region emerges adjacent to the horizon, and both the CTCs region and the ergoregion exhibit a multi-lobe structure. 

The existence of a negative-energy region (NER) is essential for the Penrose process. In the Kerr case with no magnetic field (i.e., ${\cal Q} = 0$ and $B=0$), the ergoregion serves as the  NER, where the mechanical Penrose process can occur. A non-zero ${\cal Q}$ alters the NER, leading to a multi-lob structure of NER when $|{\cal Q}|$ is sufficiently large (as depicted in Fig. \ref{NERQ-1}), similar to the structures of the ergoregion and the CTC region. In the absence of electromagnetic interaction, the NER is always situated within the ergoregion. 

However, incorporating the electromagnetic interaction drastically alters the NER, as illustrated in Figs. \ref{NERQ-2} and \ref{NERQ-3}. For $\bar{q} B >0$, the NER extends beyond the ergoregion in specific angular directions, even for small $\bar{q} B$. As $\bar{q} B$ increases beyond a threshold, the NER region completely encloses the ergoregion and reverts to a simply-connected region, implying that energy extraction can occur far from the ergoregion. Notably, when ${\cal Q}$ is sufficiently positive, the ergoregion does not intersect the equatorial plane, whereas the NER does---indicating that the pure mechanical Penrose process cannot occur in the equatorial plane, while the MPP can. Conversely, for $\bar{q} B <0$, the NER remains within the ergoregion and shrinks as $|\bar{q} B|$ increases. These results suggest that energy extraction is more favorable when $q$ and $B$ share the same sign.

After analyzing the influences of ${\cal Q}$ and $B$ on the NER, we consider a simplified scenario where the MPP occurs in the equatorial plane and derive the general formalism for the energy extraction efficiency $\eta$ (see Eq. (\ref{Efficiency-3})). Without loss of generality, we examine a specific event: the beta decay of a neutron (particle $0$) into an electron (particle $1$) and a proton (particle $2$). As illustrated in Figs. \ref{Fig: Efficiency-1}, \ref{Fig: Efficiency-2}, \ref{Fig: Efficiency-3} and \ref{Fig: Efficiency-4},  $\eta$ can reach extremely high values, far exceeding those of the mechanical Penrose process, due to the effects of ${\cal Q}$ and $B$. Specifically, the primary impacts of the two parameters on $\eta$ are as follows:
\begin{itemize}
    \item For $B=0$ (Figs. \ref{Fig: Efficiency-1} and \ref{Fig: Efficiency-2}), a more negative ${\cal Q}$ and a higher spin $a$ enhance energy extraction. The maximum efficiency $\eta_{\rm max}$ can exceed $100\%$ if ${\cal Q}$ is sufficiently negative. As ${\cal Q}$ transitions from positive to negative, $\eta_{\rm max}$ undergoes three distinct evolutionary stages: an initially slow growth, followed by a rapid increase, and finally a return to slow growth. 

    \item For $B\neq 0$ and $|B| > 10^{-18}$, $\eta$ is dominated by the electromagnetic term, with $\eta_{\rm max} \sim (0.1 - 1) \hat{\bar{q}} B \sim 10^{19-20} |B| \%$. Consequently, $\eta$ scales with $|B|$ and can reach extremely high values even for small $|B|$. Optimal energy extraction occurs under two conditions: (i) when $q$ and $B$ share the same sign, and (ii) when ${\cal Q}$ is more positive. Similar to the $B=0$ case, $\eta_{\rm max}$ undergoes three distinct evolutionary stages as ${\cal Q}$ varies from negative to positive values.
\end{itemize}

The potential astrophysical applications of the MPP have been discussed in several studies, one notable example being the explanation of the origin and production mechanisms of extragalactic ultra-high-energy cosmic rays with energies exceeding $10^{20} \ {\rm eV}$ \cite{Tursunov:2020juz,Tursunov:2022iqk}. In the specific process we consider, the escaping proton can attain an estimated maximum energy (taking $a=0.8$ as an example) of $E_{p^+} \sim c_0 \times 10^{26} |B|\ \text{eV}$, with  $c_0$ ranges from roughly $4$ to $8.5$ as ${\cal Q}$ varies from $-5$ to $2$. This energy scale aligns with the observed energies of extragalactic ultra-high-energy cosmic rays when $|B| > 10^{-6}$.

Several important issues remain open for future investigation. One interesting direction is to systematically compare MPP with other energy extraction mechanisms, such as the collisional Penrose process \cite{piran1975high}, superradiant scattering \cite{Teukolsky:1974yv}, the BZ mechanism \cite{Blandford:1977ds}, the MHD Penrose process \cite{1990ApJ...363..206T}, the BSW mechanism \cite{Banados:2009pr}, and the Comisso-Asenjo mechanism \cite{Comisso:2020ykg}. Another promising avenue is to explore alternative configurations of the magnetic field beyond the uniform case considered here. We leave these questions for future work.

\begin{acknowledgments}

	This work is supported by the National Natural Science Foundation of China (NNSFC) under Grant No 12075207. We would like to thank the anonymous referee for his/her very helpful suggestions, which have indeed improved this manuscript.

\end{acknowledgments}

\bibliographystyle{utphys}
\bibliography{ref}

\end{document}